\documentclass[%
 aip,
 amsmath,amssymb,
 reprint,%
]{revtex4-1}

\usepackage{graphicx}
\usepackage{dcolumn}
\usepackage{bm}
 
\usepackage[utf8]{inputenc}
\usepackage[T1]{fontenc}
\usepackage{mathptmx}
\usepackage{etoolbox}

\makeatletter
\def\@email#1#2{%
 \endgroup
 \patchcmd{\titleblock@produce}
  {\frontmatter@RRAPformat}
  {\frontmatter@RRAPformat{\produce@RRAP{*#1\href{mailto:#2}{#2}}}\frontmatter@RRAPformat}
  {}{}
}%
\makeatother
\begin{document}

\preprint{AIP/123-QED}

\title[]{Fast loaded dual species magneto-optical trap of cold Sodium and Potassium atoms with light-assisted inter-species interaction}

\author{Sagar Sutradhar}
\author{Anirban Misra}
\author{Gourab Pal}
\author{Sayari Majumder}
\author{Sanjukta Roy}
\author{Saptarishi Chaudhuri}%
\email{srishic@rri.res.in}
\affiliation{Raman Research Institute, C. V. Raman Avenue, Sadashivanagar, Bangalore 560080, India}

\date{\today}

\begin{abstract}
We present the design, implementation and detailed experimental characterization and comparison with  numerical simulations of two-dimensional Magneto-optical traps (MOT) of bosonic $^{23}$Na and $^{39}$K atoms for loading the cold atomic mixture in a dual-species 3DMOT with a large number of atoms. We report our various measurements pertaining to the characterisation of the two 2D$^+$MOTs via the capture rate in the 3DMOT and also present the optimised parameters for the best performance of the system of the cold atomic mixture. 
In the optimised condition, we capture more than $3 \times 10^{10}$ $^{39}$K atoms and $5.8 \times 10^8$ $^{23}$Na atoms in the 3DMOT simultaneously from the individual 2D$^+$MOTs with the capture rate of $5 \times 10^{10}$ atoms/sec and $3.5 \times 10^8$ atoms/sec for $^{39}$K and $^{23}$Na, respectively. We also demonstrate improvements of more than a factor of 5 in the capture rate into the 3DMOT from the cold atomic sources when a relatively high-power ultra-violet light is used to cause light-induced atomic desorption in the 2D$^+$MOT glass cells. A detailed study of the light assisted interspecies cold collisions between the co-trapped atoms is presented and interspecies loss coefficients have been extracted to be, $\beta_{NaK} \sim 2 \times 10^{-12}$ cm$^3$/sec.
The cold atomic mixture would be useful for further experiments on Quantum simulation with ultra-cold quantum mixtures in optical potentials.
 
\end{abstract}

\maketitle

\section{\label{sec:level1}Introduction}
Ultra-cold quantum gases in optical potentials offer a versatile platform for Quantum Simulation \cite{BlochRMP2008,Bloch2012,Gross2017Science,Schafer2020, Denschlag_2002}, precision measurements \cite{YeIJMPD2007} and Quantum Technologies \cite{Zhang2016} due to the high degree of controllability of such systems such as inter-atomic interaction, dimensionality, spin states and external potentials. This makes ultra-cold atomic ensembles an ideal `quantum toolbox' leading to unprecedented progress in this research field. 

\par
Quantum gas mixtures with dual atomic species has attracted considerable interest since they offer a wealth of novel possibilities. Quantum degenerate mixtures realized by using single atomic species in different Zeeman sub-levels \cite{MyattPRL1997,HallPRL1998,Stenger_1998}, multiple isotopes of same species or different atomic species \cite{Modugno1320,HadzibabicPRL2002,ModugnoPRL2002,RegalPRL2004,SilberPRL2005,Gunter_PRL_2006, PappPRL2008,Ospelkaus_PRL_2006,Catani_PRA_2008,TaglieberPRL2008, MacCarronPRA2011,PasquiouPRA2013,Barbutscience2014, 2015_Arlt, 2018_Grimm, Castilho_2019} can be used to investigate novel quantum phases hitherto unexplored in single atomic species. For example, the physics of impurities coupled to degenerate gas \cite{ChikkaturPRL2000, OspelkausPRL2006, BrudererELP2008} is of fundamental importance in condensed matter systems. Novel exotic quantum phases such as quantum droplets in a spin mixture of Bose gases have recently been proposed \cite{PetrovPRL2015} and observed in homo- and hetero-nuclear quantum mixtures \cite{CabreraScience2018, SemeghiniPRL2018, 2019_Chiara}.
Quantum mixtures can also be used to create hetero-nuclear stable polar molecules \cite{2010PRAHutson, VogesPRl2019, ParkPRL2015} which is useful to study controlled ultra-cold chemistry  \cite{OspelkausScience2010} as well as long-range anisotropic dipolar interactions \cite{DeMarcoScience2019, TakekoshiPRL2014,Damski_PRL_2003,Lahaye_2009}.
\par
A Quantum degenerate mixture of sodium and potassium is an attractive combination for a hetero-nuclear quantum mixture experiment. Both the Bose-Bose mixture ($^{23}$Na-$^{39}$K, $^{23}$Na-$^{41}$K) and Bose-Fermi mixture ($^{23}$Na-$^{40}$K) can be obtained opening up a myriad of possibilities for exploring the many-body physics arising due to the interplay between inter-species and intra-species interaction with quantum statistics playing a significant role. Another important advantage of the combination of sodium and potassium for the hetero-nuclear quantum mixture is that the Na-K ground-state polar molecules \cite{2022_Bloch, 2016_Zwerlein} are chemically stable as compared to other combinations of inter-species hetero-nuclear molecules with a large dipole moment of $\sim$ 2.72 Debye paving the way to explore long-range dipolar interaction for quantum simulation \cite{2008_Baranov}.
\par
In this article, we describe our experimental setup to realize an ultra-cold atomic mixture of $^{23}$Na and $^{39}$K atoms in a dual-species magneto-optical trap (3DMOT) loaded from cold atomic beams produced via two independent, compact and efficient two-dimensional magneto-optical traps (2D$^+$MOTs) of $^{23}$Na and $^{39}$K. We also present the detailed characterisation of the performance of the cold atom sources of both $^{23}$Na and $^{39}$K atoms to obtain the optimised experimental parameters for the best possible performance of the cold atomic beam sources. 

The various sections in this article are organised as follows: In section \ref{section:Expt_setup}, we provide a detailed description of the experimental system including the ultra-high vacuum assembly and laser systems. In section \ref{sec:atomic_source}, we focus on the characterisation and performance of the cold atomic beam sources. In section \ref{sec:sim} we give a detailed theoretical description of the numerical simulations performed in order to compare with the experimental results of the atomic sources.
We have provided a complete system performance study in section \ref{sec:2D_plus_experimental_results}.
 Finally, we discuss about the interspecies light-assisted collisions between hetero-nuclear cold atoms in section \ref{sec:Beta}.

\begin{figure*}[ht]
\centering
\includegraphics[scale=0.9]{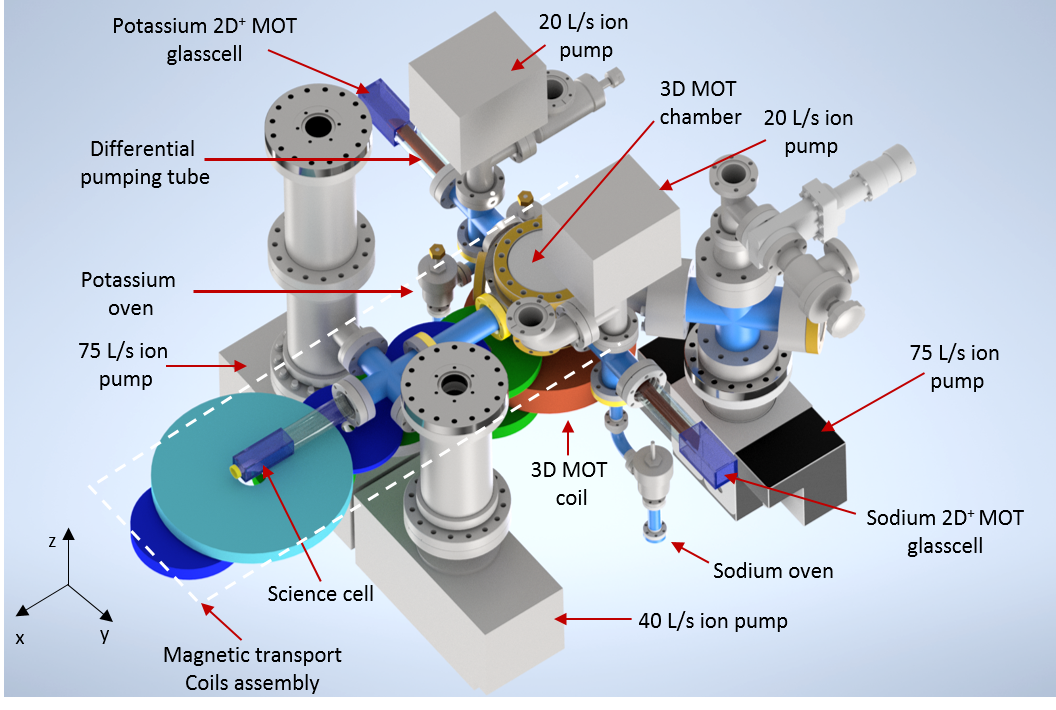}
\caption{(Color online) A schematic of the vacuum assembly. The two-species MOT is loaded from two independent  $2D^{+}$MOTs as sources of cold $^{23}$Na and $^{39}$K atoms.  The dual-species 3DMOT is produced in a spherical octagonal chamber.  The UHV side is pumped by three large-capacity ion pumps whereas the two independent source regions are pumped with two 20 l/s ion pumps. Coils made of hollow-cored water-cooled copper tubes placed outside 3DMOT chamber are used to generate the quadrupole magnetic field for trapping of atoms. A single-arm magnetic transport allows transferring the cloud to the `science cell' with large optical access.}
\label{fig:Vacuum_setup}
\end{figure*}

\section{Experimental Setup}
\label{section:Expt_setup}
\par In this experimental setup, a large number of $^{23}$Na and $^{39}$K atoms are simultaneously captured in a dual-species 3DMOT from two independent sources of the cold atomic beams. There are stringent requirements on the design of the apparatus such as good optical access for trapping laser beams as well as detection, ultra-high vacuum to ensure longer trap lifetime of the atoms and high magnetic field gradient for magnetic trapping. Our experimental setup is designed and built up to fulfil these requirements and enable further experiments on the quantum degenerate mixture in both magnetic and optical potentials. 
\par
The conflicting requirements of having a large number of atoms for experiments on degenerate quantum gases as well as a long lifetime of the atomic cloud has led to the design of multi-chamber vacuum systems for such experiments where the MOT is loaded from a cold atomic beam source instead of the background vapour. Examples of such cold atomic beam sources are: Zeeman slower \cite{PhillipsPRL1982}, Low velocity intense source \cite{1996PRLLu}, 2DMOT \cite{2002PRASchoser}, 2D$^{+}$ MOT \cite{2006_Chaudhuri,1998PRADieckmann}, and pyramidal MOTs \cite{2021OERavenhall}.
 Amongst such possibilities, $2D^{+}$MOT offers the most compact design with the most efficient performance. For $^{23}$Na and $^{39}$K atoms, Zeeman slowers \cite {PhillipsPRL1982,Na_Zeemanslower_Straten,Pedrozo_Pe_afiel_2016}  and 2DMOTs\cite{CataniPRA2006,GrobnerJMO2016,Ferrari,Pedrozo_Pe_afiel_2016}, have been realized. In the case of hetero-nuclear atomic species mixture, to the best of our knowledge, our experiment is the first demonstration where both the atomic species are simultaneously derived from compact $2D^{+}$MOT configurations. 

\subsection{Vacuum assembly} \label{section:UHV}

A schematic view of our vacuum system is shown in Fig.\ref{fig:Vacuum_setup}. A spherical octagon-shaped chamber for 3DMOT, made with non-magnetic stainless steel (Kimball physics-MCF600-SphOct-F2C8) placed at the centre of the vacuum manifold is attached with two independent 2D$^+$MOT glass cells (Precision Glassblowing, Colorado, USA). 
For both $^{23}$Na and  $^{39}$K atoms, the vacuum chamber of the 2D$^+$MOT consists of a cuboidal glass cell (dimensions 85\,mm $\times$ 40\, mm $\times$ 40\, mm), whose longitudinal axis is aligned horizontally and placed along the axis of a differential pumping tube connecting the 2D$^+$MOT glass cell and the 3DMOT chamber. The atomic beam is prepared along the longitudinal axis of the glass cell. The differential pumping tube was made from a single block of oxygen-free highly conductive (OFHC) copper. One end of the tube is a $45^{\circ}$-angled polished mirror with a round surface of diameter 18\, mm and placed inside the glass cell. The other end of the tube has a disk shape of diameter $\approx$ 48 mm and a thickness of 10 mm. This disk acts as a gasket between the two CF40 flanges of the  2D$^+$MOT and the 3DMOT chamber. The  $45^{\circ}$ surface of the copper tube allows the alignment of the longitudinal cooling laser beams as described later in this article.

The differential tube has a hole which originates at the center of the $45^{\circ}$ surface and runs along the axis of the tube and ends up at the UHV side of the 3DMOT chamber.  The differential pumping hole starts with a diameter of 2\, mm and then widens up in two steps over a total distance of 270\,mm. The hole reaches a diameter of 8 \,mm (6 \,mm) after the first 20 \,mm length and subsequently widens up to 14\,mm (12 \,mm) after the next 120 mm length for $^{23}$Na ($^{39}$K) tubes. 

\begin{figure}[h!]
\centering
\includegraphics[width=8.7cm]{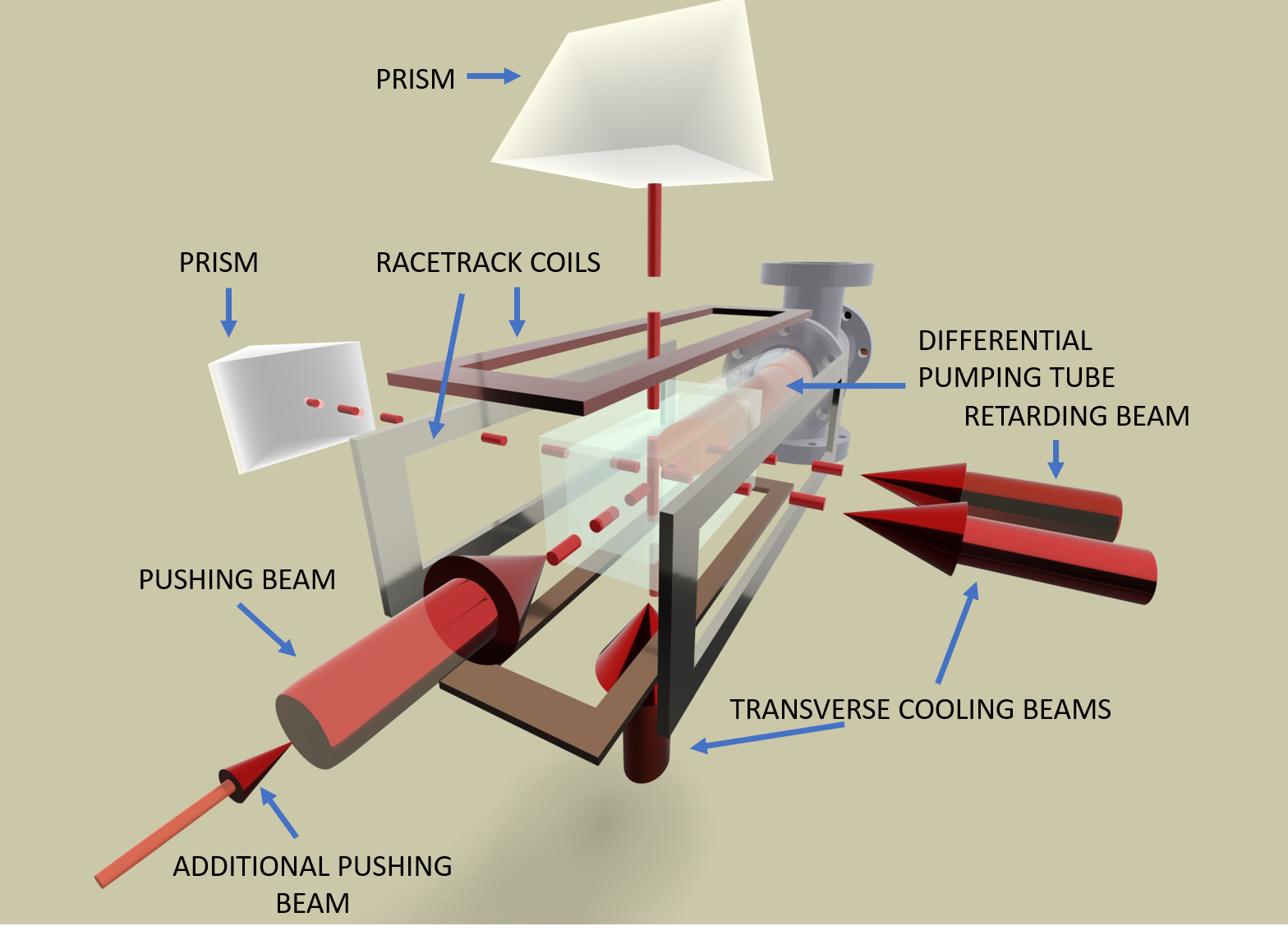}
\caption{(Color online) Schematic diagram of the $2D^{+}$MOT. Two transverse cooling beams are retro-reflected using two helicity-preserving right-angled prisms. In addition, a pair of longitudinal cooling beams (pushing and retarding beams) are aligned along the line of zero magnetic field created due to the configuration of the four race-track-shaped coils. The copper tube with a differential pumping hole connecting the 2D$^+$MOT and the 3DMOT sides is cut at an angle of 45$^o$ and mirror-polished to facilitate the passage of the retarding beam. An additional pushing beam is used to direct the cold atomic beam to the 3DMOT chamber through the differential pumping hole.}
\label{fig:2DMOTimage}
\end{figure}
 The differential pumping tube has a conductance of  0.043 l/s (0.038 l/s) for the $^{23}$Na ($^{39}$K) side. The two 2D$^+$MOT glass cells are individually pumped using two 20 l/s Ion pumps. The 3D-MOT chamber is pumped by a 75 l/s Ion pump, and the generated pressure ratio between the two chambers is 1200 (1400) for $^{23}$Na ($^{39}$K) side. 
 \par
 Additionally, our experimental system includes a magnetic transport tube and a glass cell (`science cell') of dimension 85 mm $\times$ 30 mm $\times$ 30 mm pumped by two more Ion pumps with 40 l/s and 75 l/s pumping speeds. We also occasionally use a Titanium Sublimation pump to maintain the base pressure below $10^{-11}$ mbar near the `science cell'. The base pressure near the 3DMOT chamber is measured using an ionisation gauge to be $\sim 7 \times 10^{-11}$ mbar which is also consistent with our observed cold atom trap lifetime of $\sim$ 48 s. On the other hand, both the 2D$^+$MOT glass cells are maintained at a base pressure below $10^{-9}$ mbar. 
 \par
 We have used a natural abundance source (ingot) of sodium ( Sigma Aldrich(262714-5G)). The ingot is placed inside a CF16 full nipple and attached to the glass cell through a CF16 angle gate valve (MDC vacuum). Heatings tapes are wrapped around the full nipple and the gate valve in such a way that, we could maintain a temperature gradient from the oven towards the glass cell, which ensures the sodium drifts into the cell and remains there. The purpose of the gate valve is two-fold, first, it determines the amount of flow of sodium vapour into the glass cell and, second, during replenishment of the source it would allow us to isolate the oven from the rest of the vacuum system.
\par 
 We have also used a natural abundance source (ingot) of potassium from Sigma Aldrich (244856-5G) as the source for loading atoms in the $^{39}$K 2D$^+$MOT. The design of the potassium oven is similar to the sodium one. Here we have kept natural abundance potassium and enriched $^{40}$K (10$\%$ enrichment, from Precision Glassblowing, USA), inside two different CF-16 full nipples, followed by respective CF-16 angle gate valves. These two ovens are connected and integrated with the 2D$^+$MOT glass cell.
\subsection{Laser systems}
\label{section:lasersystem}

The cooling and repumping beams for the laser cooling of sodium atoms were derived from a frequency-doubled Diode laser system (Toptica TA-SHG pro) which typically gives a total output power of 1100 mW at 589 nm ($^{23}$Na D2 transition). The laser beam from the TA-SHG pro is divided into several beams. A low-power beam (typically 5 mW) is fed into an AOM (AA optics, centre frequency 110 MHz) double-pass assembly and subsequently directed into the saturation absorption spectroscopy (SAS) setup. The spectroscopy for sodium is realized using a vapour cell of length 75\, cm from Triad technologies (TT-NA-75-V-P),  which is heated to 150$^\circ$C to create a sufficiently high vapour pressure for absorption.

The cooling beams for the 2D$^{+}$MOT as well as the 3DMOT are generated using two independent AOM (Isomet 110 MHz) double-pass setups and tuned appropriately red-detuned from the $ 3^{2}S_{1/2} |F=2\rangle \rightarrow 3 ^{2}P_{3/2}|F'=3\rangle$ transition. The repumping beams are tuned in resonance with the transition $ 3^{2}S_{1/2} |F=1\rangle \rightarrow 3 ^{2}P_{3/2}|F'=2\rangle$, by passing the cooling beams through two independent  Electro-optic modulators (EOM) (QuBig-EO-Na1.7M3). The EOMs are powered by two independent drivers (QuBig-E3.93KC), and each sideband has typically 20\% of the power of the carrier (cooling) frequency.  The co-propagating cooling and repumping beams are injected into their respective polarization-maintaining (PM) fibers and transferred to the experimental optical table for the realization of the 2D$^+$MOT and the 3DMOT. 


For potassium atoms, we use two independent External Cavity Diode Lasers (ECDL) from Toptica Photonics for deriving the cooling (DL pro) and repumping (DL 100)  laser beams. Each of these laser outputs is amplified using two independent tapered amplifiers (Toptica BoosTA pro) with a maximum output power reaching 2 W. The output of each of the Potassium lasers is divided into two beams, the one with low power $\approx$ 5mW is fed into the SAS setup. The spectroscopy is realized with a glass vapour cell of length 5$\,$cm, in which a K-sample with natural abundance is heated to 50$^\circ$C. The other output beams from the two Potassium lasers are injected into the Tapered Amplifiers (TA). The amplified output beams of the TAs are split into several beams and sent through the corresponding AOM ( AA Optics, 200 MHz) double pass configurations to prepare the beams at the appropriate frequencies to be used as the cooling and repumping beams for the 2D$^{+}$MOT and the 3DMOT. 
The K- cooling laser is offset-locked to the $4S_{1/2}|F=2\rangle \rightarrow 4P_{3/2} |F'=3\rangle$ transition of $^{39}$K atoms, while the repumping laser is locked to the  $4S_{1/2}(|F=1\rangle, |F=2 \rangle) \rightarrow 4P_{3/2}$ crossover transition. 
\par
The laser beams transferred to the main experimental table using PM fibers (Schafter-Kirchhoff GmbH) are out-coupled by the corresponding fiber-collimators, which provides a collimated Gaussian beam of $1/e^2$ diameter of 12 mm.

\section{Atomic beam Source} \label{sec:atomic_source}
The 3DMOT can be loaded efficiently from a cold atomic beam with a high capture rate of atoms. 
The preparation of an atomic beam requires high atomic vapour pressure (in the range, of 10$^{-8}$-10$^{-6}$ mbar). Generally, it is prepared in a section spatially separated from the 3DMOT section (which requires a UHV environment). The atomic beam is directed through a differential pumping tube into the 3DMOT section, which maintains a pressure difference of a few orders of magnitude between the 2D$^{+}$MOT region and the UHV 3DMOT region. This atom loading scheme not only keeps the  3DMOT in a UHV range to minimize the collisions with the room-temperature atoms but also loads the 3DMOT with a large number of atoms.  This creates a favourable starting point for proceeding towards evaporative cooling of the cold atomic mixture to quantum degeneracy.

We have employed two independent and spatially separated 2D$^{+}$MOTs as the cold atomic beam sources for $^{23}$Na and  $^{39}$K, which provide cold collimated atomic beams to load the 3DMOT in the UHV chamber through two opposite ports. 

\subsection{2D$^+$MOT} \label{sec:2D_plus_MOT}

The 2D$^+$MOT is created by two orthogonal retro-reflected elliptical (circular) laser beams in presence of a two-dimensional quadrupole magnetic field for $^{39}$K ($^{23}$Na) atoms. The elliptical beams are chosen slightly convergent with  $1/e^2$ diameter of incident beams to be 36 mm (24 mm) along the atomic beam direction and normal to the atomic beam axis is 24 mm (24 mm) at the 2D$^+$MOT cloud position for $^{39}$K ($^{23}$Na) atoms. The circularly polarised transverse cooling beams are retro-reflected using right-angled prisms which preserves the helicity of the beams via two total internal reflections. The degree of convergence has been chosen to accommodate the reflection of the uncoated glass cell surfaces (typically, each surface has 4 per cent reflectance), such that we obtain the same intensities of the incident and the retro-reflected beam at the position of the atomic cloud. The 2D quadrupole magnetic field is realized by two pairs of racetrack coils in an anti-Helmholtz configuration which creates a line of zero magnetic fields along the centre of the magnetic coil configuration. The atoms cooled in the transverse direction are confined around the zero magnetic field line of the 2D quadrupole magnetic field.

The performance of the $2D^{+}$ MOT is enhanced by integrating a pair of counter-propagating laser beams with Gaussian width of 12 mm along the atomic beam direction which forms a longitudinal optical molasses cooling configuration. In this configuration, the optical molasses reduces the longitudinal velocities which allow the atoms to spend more time in the transverse cooling region.  This reduces the transverse velocities of the atoms thereby reducing the divergence of the cold atomic beam. As a result, the atoms go through the differential pumping tube without much loss of atoms thereby increasing the cold atomic beam flux loading the 3DMOT. The longitudinal cooling beams are referred to as the pushing and the retarding beams as shown in Fig.\ref{fig:2DMOTimage}. The retarding beam is aligned counter-propagating to the direction of the cold atomic beam. The longitudinal cooling using optical molasses reduces the longitudinal velocity of the cold atomic beam within the capture velocity of the 3DMOT in the UHV chamber thereby increasing the atoms captured in the 3DMOT. The retarding beam has a dark cylindrical region due to the hole in the mirror overlapping with the pushing beam in the counter-propagating direction which creates an imbalance of radiation pressure along the shadow region and helps in pushing the cold atomic beam to the UHV chamber through the differential pumping hole. An additional pushing beam with a Gaussian width of 1.15 mm (1.3 mm) for  $^{23}$Na ($^{39}$K) was aligned along with the pushing and retarding beams which pushes the atomic cloud into the 3DMOT chamber.


Two pairs of race-track-shaped magnetic coils in the anti-Helmholtz configuration are placed around the 2D MOT glass cell symmetrically. The transverse magnetic field gradient is 21 G/cm/A (12 G/cm/A) for  $^{23}$Na ($^{39}$K).
\par
We use LabVIEW interfaced PXIe system (NI PXIe 1062Q chassis) containing a digital card (NI 6535) and an analog card (NI 6538), to precisely control the intensity as well as the detuning of the cooling and repumping laser beams through AOMs. 
Also, all the trigger lines (required for the camera trigger, RF switches and IGBT gate trigger for magnetic field switching) are drawn from the digital channels which have 100\,ns time resolution. 
Acquisition of experimental images is done through Thorlabs scientific camera (Thorlabs CS2100M) of high quantum efficiency (61\% at 600nm), integrated with the LabVIEW.


\section{Numerical simulation}
\label{sec:sim}

\begin{figure}[h]
    \centering
    \includegraphics[width=0.35\textwidth]{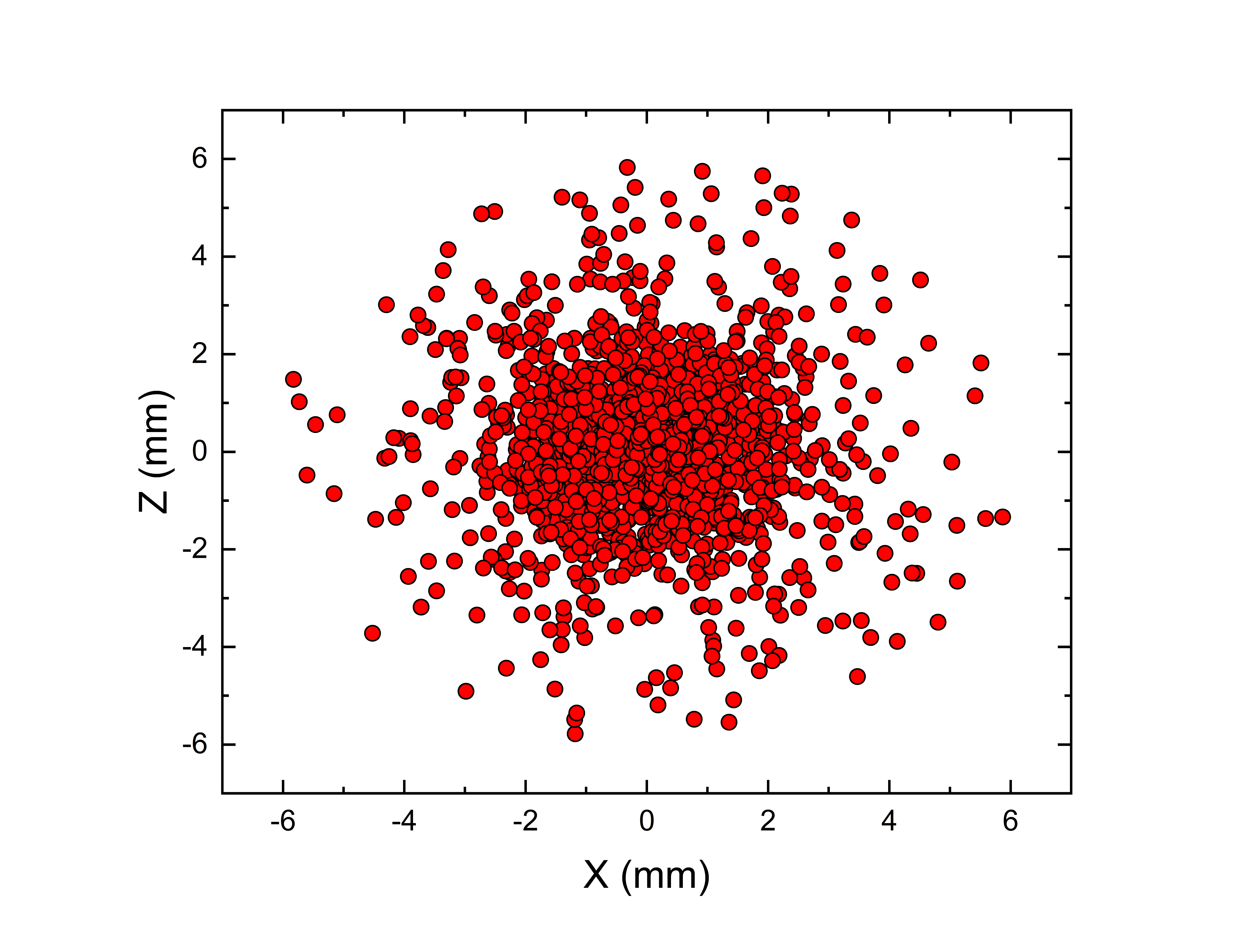}
    \caption{The final position of the particles on the transverse plane (i.e. XZ plane) at the end of their trajectories, at the 3DMOT position.}
    \label{figure:tpm}
\end{figure}

\begin{figure}[h]
    \centering
    \includegraphics[width=0.35\textwidth]{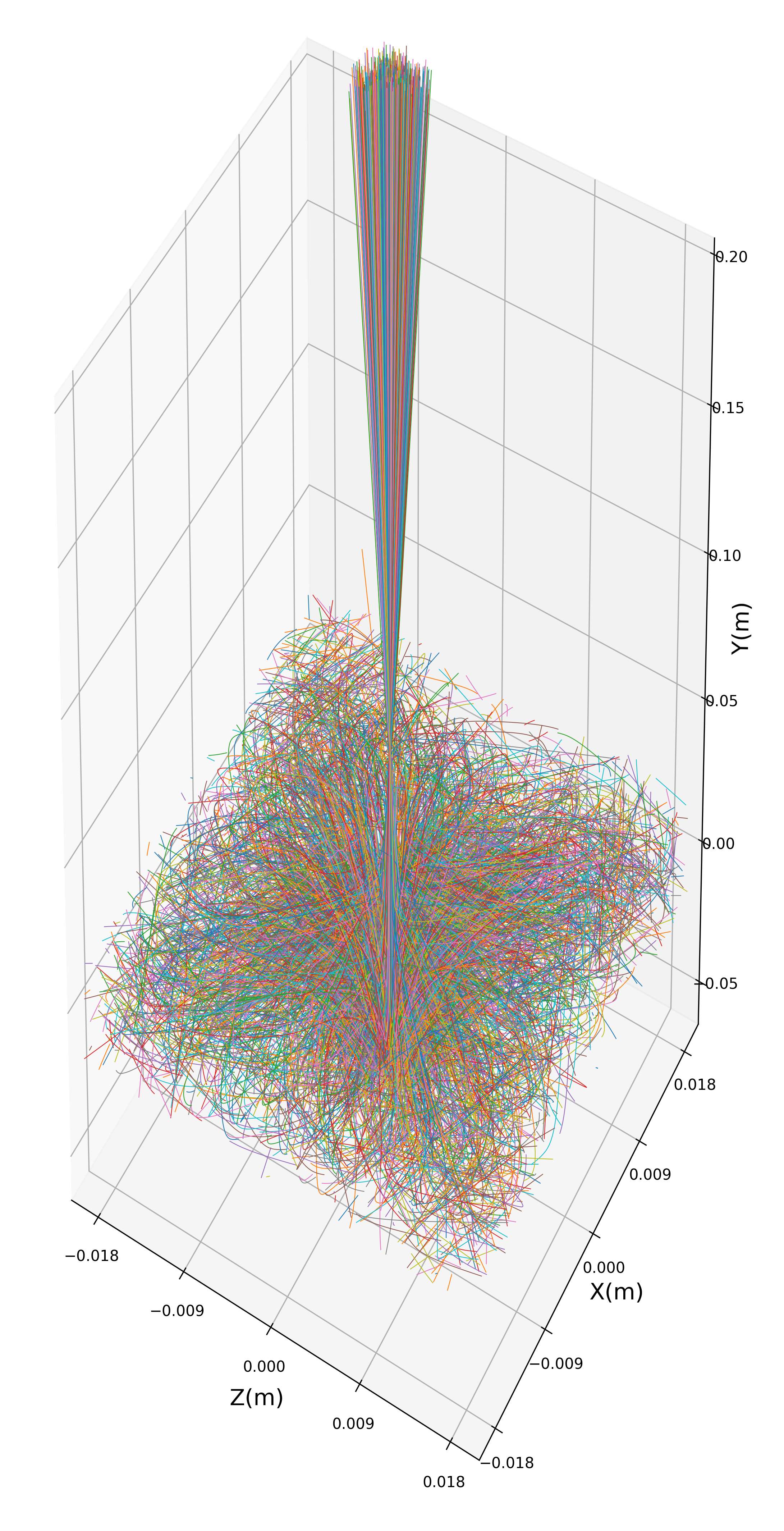}
    \caption{Trajectory of all the particles coming out of the 2D$^+$MOT glass cell.}
    \label{figure:3D_trajectories}
\end{figure}

A numerical simulation has been performed to model the characteristics of the 2D$^+$MOT as an atom source to load atoms in the 3DMOT in the UHV. In this simulation, the trajectory of each particle captured from background vapour is calculated using their equations of motion. The initial position of the atoms is chosen randomly within the 2D$^+$MOT glass cell excluding the portion containing the copper tube's end that protrudes into the glass cell. The velocities of the atoms are chosen according to Maxwell-Boltzmann distribution at a certain temperature $T$ (in Kelvin) using the Monte Carlo method. It is assumed that all the particles we will consider for mapping their trajectories have velocities within the capture velocity of that atomic species in the 2D$^+$MOT at that particular temperature. The capture velocity is determined by the temperature of the atoms as well as the intensity, detuning and size of the cooling beams of the 2D$^+$MOT.
After assigning an initial position and velocity to each particle, each of their trajectories is mapped out using RK4 (Runge-Kutta 4) method in the presence of radiation force due to the cooling laser beams, and the magnetic field gradient in the 2D$^+$MOT. While calculating the force, a simplified model of two-level atoms was assumed where the atoms are subjected to the cooling beams with frequency red-detuned to the cooling transition. The velocity-dependent force on the atoms is imparted by the four transverse cooling laser beams and a pair of longitudinal cooling beams along the line of zero magnetic field axis within the 2D$^+$MOT glass cell along the longitudinal Y-direction. The magneto-optical trapping happens only in the transverse directions determined by the intensity and polarization of the transverse cooling beams as well as the corresponding magnetic-field gradient in the transverse (XZ) directions.\\

For each axis, the total force on each atom in the 2D$^+$MOT glass cell is contributed from two directions denoted by `$+$' and the `$-$' directions of a particular axis. The detuning of the cooling laser beams plays a very important role in determining the force on the atoms. The effective detuning $\delta_{\pm}$ of the beams are given by 
\begin{equation}
\delta_{\pm} = \delta \pm k.v \mp \mu_{eg} B(r)/\hbar 
\end{equation}
where $\delta$ is the detuning of the laser beam from the atomic resonance. The total magneto-optical force on the atoms is given by $F = F_+ + F_-$ where 
\begin{equation}
F_{\pm}=\pm \frac{\hbar k \Gamma}{2}\frac{s_{0}}{1+s_{0}+\left(2\delta_{\pm}/\Gamma\right)^{2}}
\end{equation}

where , $k$ is the wave vector of the laser beams, $\hbar$ is the Planck's constant and $\Gamma$ is the natural linewidth of the cooling transition, $v$ is the velocity of the atoms, $\mu_{eg}$ is effective magnetic moment for the cooling transition, $B(r)$ is the magnetic field along the particular axis, $s_0$ is the saturation parameter given by $s_{0} = I/I_{sat}$ where $I$ is the intensity of the cooling beams and $I_{sat}$ is the saturation intensity of the atomic transition. Along the longitudinal $Z$ axis, $B(r)=0$ and for the transverse axes(i.e. $X$ or $Y$), $B(r)=\sqrt{B_{x}^2+B_{y}^2}$. The `$+$' and the `$-$' signs in the force equations are chosen according to the direction in which the force is calculated on the atoms with respect to the centre of the 2D$^+$MOT glass cell.\\
\begin{equation}
\mu_{eg}=(g_{e}m_{Fe}-g_{g}m_{Fg})\mu_{B}
\end{equation}
 where $g_{e}m_{Fe}$ ($g_{g}m_{Fg}$) is calculated for the excited state (ground state), and $\mu_{B}$ is the Bohr magneton. The value of $(g_{e}m_{Fe}-g_{g}m_{Fg})$ has been obtained to be 1 for the $D_{2}$ cooling transition of both $^{23} Na$ and $^{39} K$ atoms and hence $\mu_{eg}=\mu_{B}$ for both the atomic species. 

\subsection*{Simulation results}

In this section,  we present the results obtained from the numerical simulation. The final positions of the particles at the end of their trajectories obtained from the numerical simulation are shown in Fig.~\ref{figure:tpm}. The trajectories of all the particles coming out of the 2D$^+$MOT glass cell is shown in Fig.~\ref{figure:3D_trajectories}. The results of the Numerical simulation for the capture rate of the $^{39}$K-MOT and the $^{23}$Na-MOT as a function of the corresponding 2D$^+$MOT cooling intensity per beam are compared with the corresponding experimental results as shown in Fig.~\ref{figure:Fig5}. Since, we have performed the simulation considering typically 10$^5$ atom trajectories in the cooling volume, whereas the total number of atoms in 2D$^+$MOT cooling volume is governed by the partial pressure, an overall scaling factor is used accordingly. Using the same numerical technique, we have also compared the 2D$^+$MOT flux as a function of magnetic field gradient with the experimental measurements as shown in Fig.~\ref{figure:Fig8}.

\section{Experimental results}

\label{sec:2D_plus_experimental_results}
For our experiment, the essential parameters which characterise the performance of the two 2D$^+$MOTs are the loading rates into the corresponding 3DMOT for $^{23}$Na and  $^{39}$K atoms. We experimentally studied its dependence on several 2D$^+$MOT parameters, such as the vapour pressure in the 2D$^+$MOT glass cell, the total cooling beam intensities, 2D$^+$MOT magnetic field gradient, the detuning of the cooling and repumping beams, intensity ratios between the repumping and cooling beams as well as the pushing and retarding beams. The optimized values of these parameters are displayed in Tab.~\ref{tab:2DMOTKparameters}. Additionally, we have also observed a significant enhancement in the performance of the 2D$^+$MOT for both the atomic species when we use Light Induced Atomic Desorption (LIAD) \cite{2010_Chandra_Raman, 2015_Campo} in both the 2D$^+$MOT vacuum manifolds. 
\par
\begin{figure}[ht]
\centering
\includegraphics[width=8.8cm]{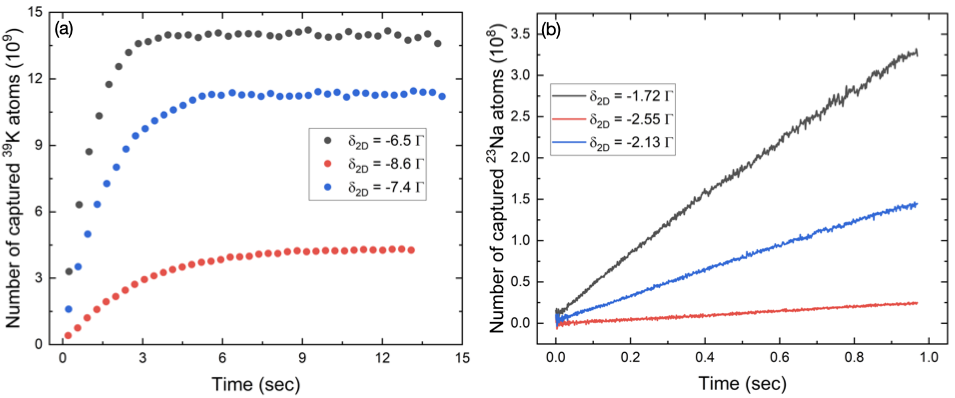}
\caption{ Number of atoms captured in the (a) $^{39}$K-MOT (b) $^{23}$Na-MOT as a function of time for various detunings of the cooling beam of the corresponding 2D$^{+}$ MOT.}
\label{figure:Fig3}
\end{figure}

We determine the capture rate of atoms into the 3DMOT using fluorescence measurements. We present our typical measurements from $^{39}$K ($^{23}$Na) 3DMOT using fluorescence images recorded on a CCD camera in Fig.~\ref{figure:Fig3}(a)(Femtowatt detector in Fig.~\ref{figure:Fig3}(b)). The number of atoms in the 3DMOT as a function of the loading time is calculated using the expression for the scattering rate, where the experimental parameters are saturation intensity \cite{2018_Smith, 1995_Walker} and laser detuning (calibrated using a weak probe beam to determine the exact resonance frequency). The number of atoms captured in the $^{39}$K-MOT and $^{23}$Na-MOT as a function of time for various detunings of the cooling beam of the corresponding 2D$^{+}$ MOT is shown in Fig.~\ref{figure:Fig3}(a) and Fig.~\ref{figure:Fig3}(b) respectively. For optimised parameters, we observe a fast loading of $5\times10^{10}$  atoms in 800 ms for  $^{39}$K atoms. In the case of $^{23}$Na atoms, we observe the loading of $5\times10^{8}$ atoms in 1.2 s limited only by the two-body collisional loss rate in the bright $^{23}$Na 3DMOT.


\begin{figure}[ht]
\centering
\includegraphics[width=8.8cm]{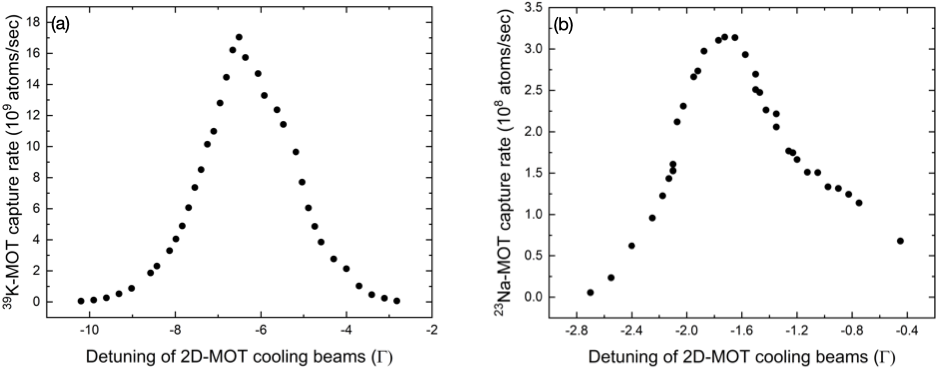}
\caption{Capture rate of (a) $^{39}$K-MOT (b) $^{23}$Na-MOT as a function of the detuning of the corresponding 2D$^{+}$ MOT cooling beams. The 3DMOT cooling beam detuning was kept fixed at -6.8$\Gamma$ and -1.4$\Gamma$ for  $^{39}$K and $^{23}$Na atoms respectively.}
\label{figure:Fig4}
\end{figure}

\begin{figure}[ht]
\centering
\includegraphics[width=8.8cm]{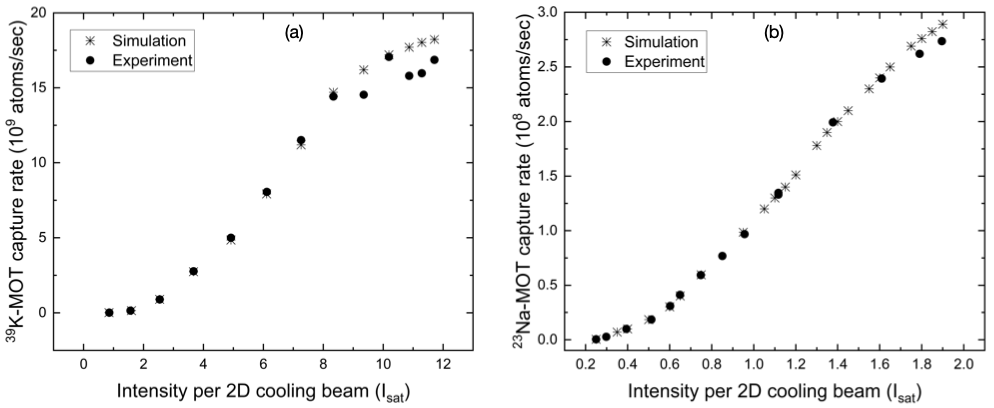}
\caption{Experimental measurements and comparison with Numerical simulation of the capture rate of (a) $^{39}$K-MOT (b) $^{23}$Na-MOT as a function of the corresponding 2D$^{+}$ MOT cooling intensity per beam. The intensity ratios between $2D^{+}$ MOT repumping and cooling beams were maintained at 0.75 and 0.18 for $^{39}$K and $^{23}$Na atoms, respectively.}
\label{figure:Fig5}
\end{figure}

\begin{figure}[ht]
\centering
\includegraphics[width=8.8cm]{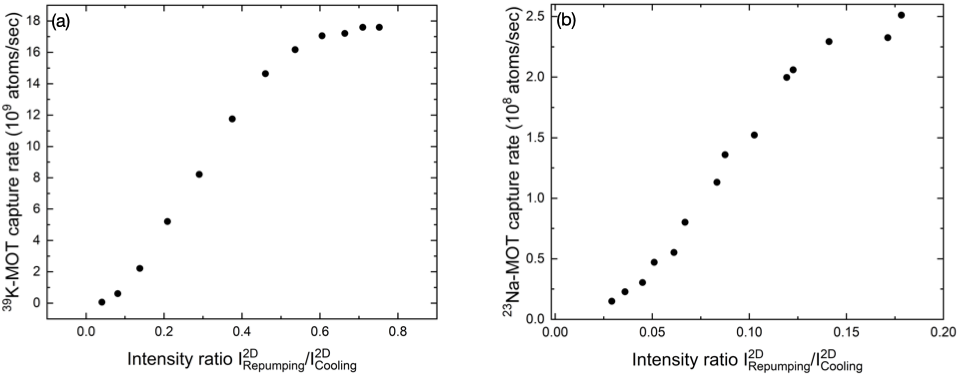}
\caption{Capture rate of (a) $^{39}$K-MOT (b) $^{23}$Na-MOT as a function of the intensity ratio between the corresponding $2D^{+}$ MOT repumping and cooling beams. In the case of $^{39}$K atoms a relatively large repumping to cooling intensity ratio of around 0.75 is required for the optimised operation of the  $2D^{+}$MOT due to the narrow spacing of the excited state hyperfine splitting. On the other hand, $^{23}$Na source works well with a relatively low repumping to cooling intensity ratio of around 0.18. 
}
\label{figure:Fig6}
\end{figure}

\begin{figure}[ht]
\centering
\includegraphics[width=8.8cm]{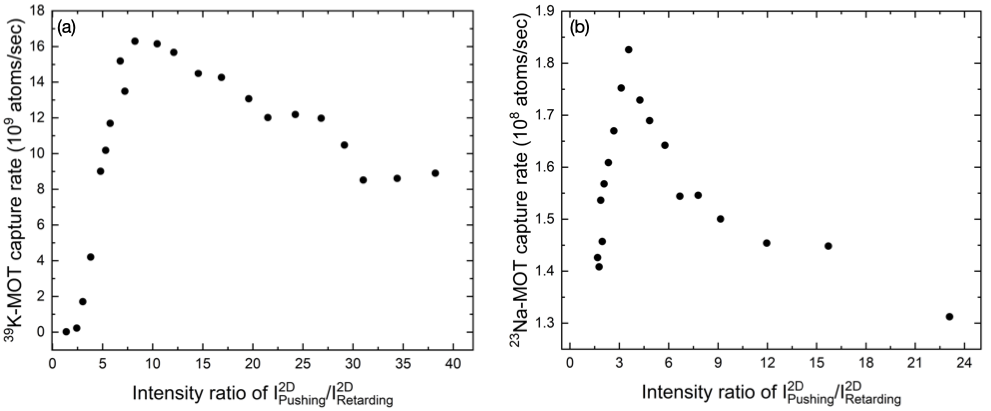}
\caption{Capture rate of (a) $^{39}$K-MOT (b)  $^{23}$Na-MOT as a function of the intensity ratio of the pushing and retarding beams of the corresponding $2D^{+}$MOT. The optimum value of the intensity ratio is experimentally obtained at 8.1 (3.6) for $^{39}$K ($^{23}$Na) atoms. the data presented in this graph is recorded at a reduced oven temperature whereas we have experimentally checked that the optimum intensity ratio remains the same as a function of the partial vapour pressure for both  $^{39}$K and $^{23}$Na atoms.}
\label{figure:Fig7}
\end{figure}

\begin{figure}[ht]
\centering
\includegraphics[width=9cm]{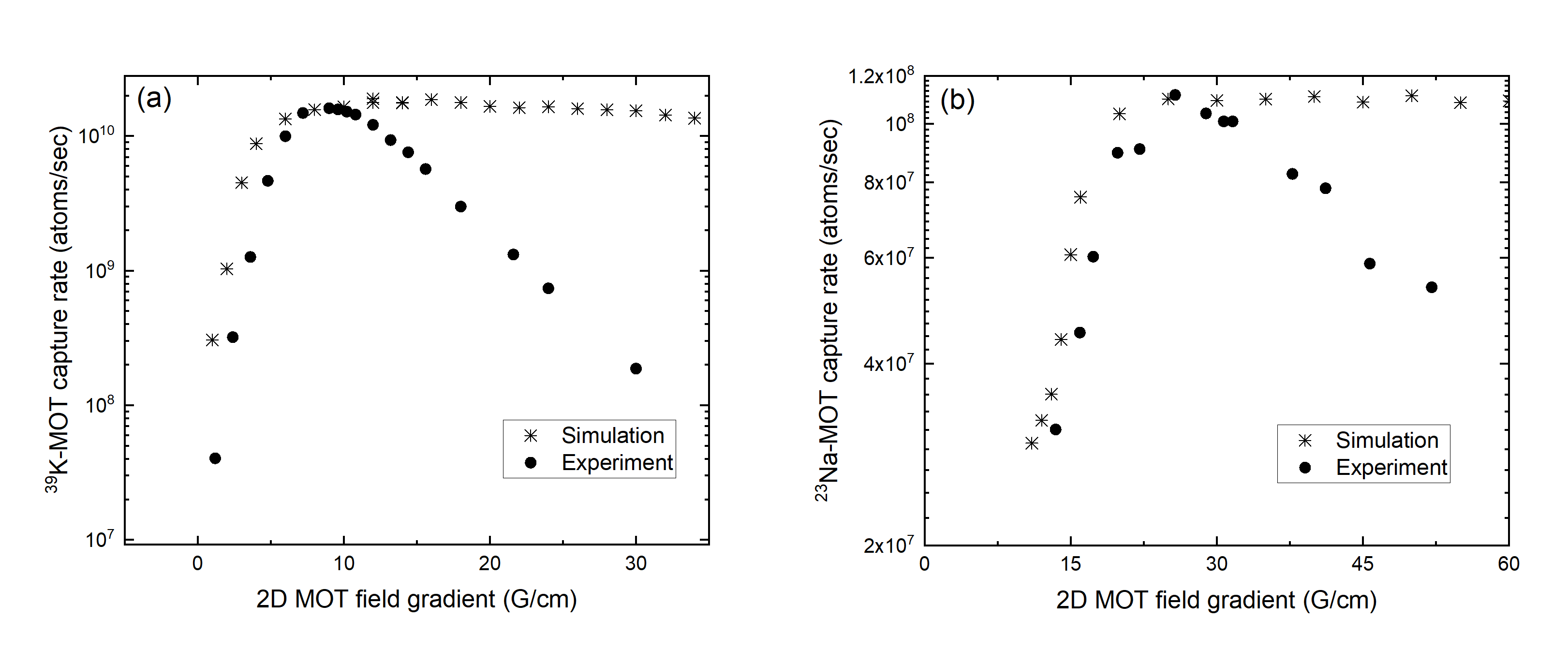}
\caption{Experimental results and comparison with the numerical simulation for the capture rate of (a) $^{39}$K-MOT (b)  $^{23}$Na-MOT as a function of the magnetic field gradient of the corresponding 2D$^{+}$ MOT. At the low magnetic field gradient the experimental data matches well with the model developed with simple two-level atom approximation. However, at high field gradients, expectedly, the results deviate especially for Potassium atoms where the excited state splitting is less compared to Sodium atoms, as elaborate in the text.}
\label{figure:Fig8}
\end{figure}

\begin{figure}[ht]
\centering
\includegraphics[width=8.8cm]{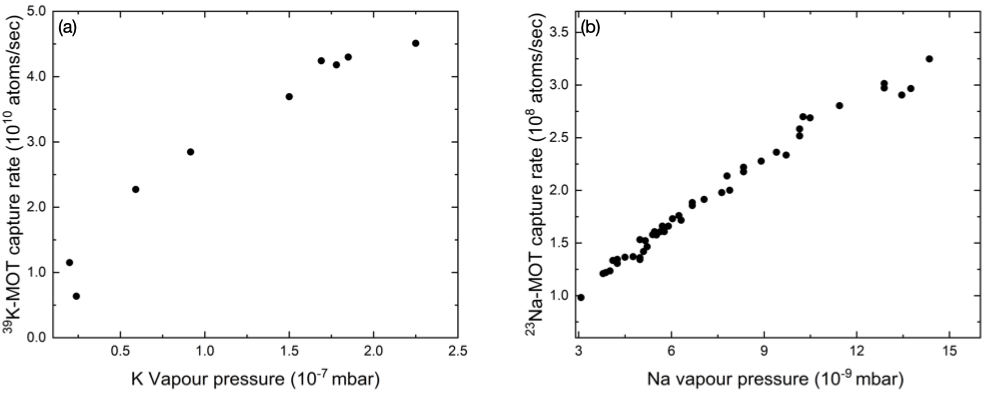}
\caption{ Capture rate of (a) $^{39}$K-MOT (b) $^{23}$Na-MOT as a function of the vapour pressure of $^{39}$K and $^{23}$Na in their respective 2D$^{+}$MOT. We observe a monotonic increase of the 3DMOT capture rates as a function of the 2D$^{+}$MOT vapor pressure for both species. For $^{39}$K, the data indicates a saturation of the capture rate above $2\times 10^{-7}$ mbar partial pressure whereas the capture rate for $^{23}$Na is only limited by the available partial vapour pressure.}
\label{figure:Fig9}
\end{figure}

\begin{figure}[ht]
\centering
\includegraphics[width=8.5cm]{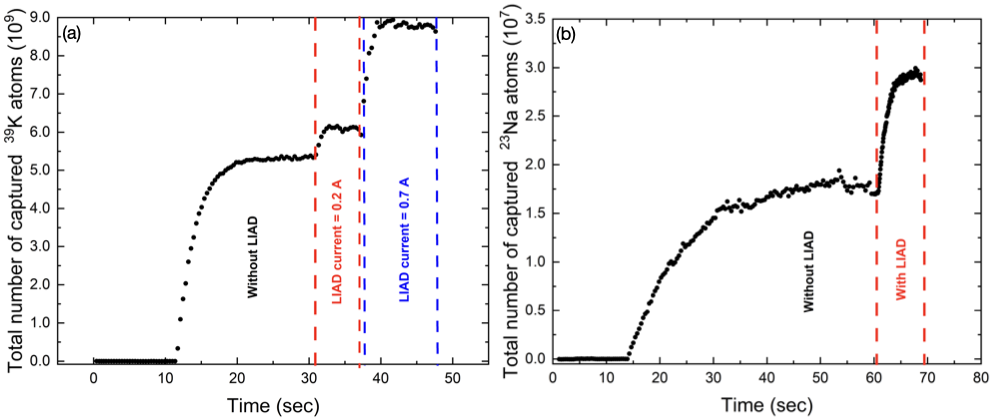}
\caption{ Total number of atoms captured in the (a) $^{39}$K-MOT (b) $^{23}$Na-MOT as a function of time demonstrating the effect of switching on the LIAD (Light-induced atomic desorption) for the corresponding 2D$^{+}$ MOT.}
\label{figure:Fig10}
\end{figure}


Fig.~\ref{figure:Fig4} shows the dependence of the capture rate of the $^{39}$K-MOT and the $^{23}$Na-MOT on the detuning of the corresponding 2D$^{+}$MOT cooling beams. The curve has a maximum at detunings -6.5$\Gamma$ and -1.8$\Gamma$ for $^{39}$K and $^{23}$Na atoms respectively. The maximum capture rate in the 3DMOT as a function of the detuning of the 2D$^{+}$MOT cooling beams is the result of two opposing effects: the scattering force of the 2D$^{+}$MOT beams decreases with increasing detuning implying a less efficient transverse cooling for higher detuning whereas the capture velocity increases with higher detuning which increases the atomic beam flux ~\cite{metcalf1999laser}. The detuning at which the maximum capture rate in the 3DMOT is obtained represents the detuning for the most efficient transverse cooling of the atoms which produces a cold atomic beam with high flux and narrow velocity distribution so as to pass through the differential pumping hole with minimum divergence.

An additional factor affecting the behaviour of the plot in Fig.~\ref{figure:Fig4} is that we measure the $^{39}$K-MOT and $^{23}$Na-MOT capture rate rather than the 2D$^+$MOT atomic beam flux. The scattering force of the pushing beam, depends on the cooling beam detuning which is one of the factors that determines the mean velocity of the atomic beam; as a consequence, it has an effect on both the atomic flux \cite{CataniPRA2006}, as well as the loading rate of the MOT\cite{Pedrozo_Pe_afiel_2016}. 

\begin{table}[h!]
\centering

\setlength{\tabcolsep}{6pt} 
\renewcommand{\arraystretch}{1.5}

\begin{tabular}{c c c c} 
 \hline 
 2D$^{+}$MOT parameters & $^{23}$Na & $^{39}$K \\ [0.5ex] 
 \hline\hline
 $\delta^{2D}_{cooling}$ ($\Gamma$)& -1.8 & -6.5 \\ 
 $I^{2D}_{cooling}$ per beam ($I_{S}$) & 7 & 10 \\
 $I^{2D}_{repumping}/I^{2D}_{cooling}$ & 0.18 & 0.75 \\ 
 $I^{2D}_{pushing}/I^{2D}_{retarding}$ & 3.6 & 8.1 \\
 $\partial_{x}B$, $\partial_{z}B$ (G/cm) & 26 & 9 \\
 $I_{additional-push}$ ($I_{S}$) & 7.7 & 5.38 \\ 
 Vapour pressure (mbar) & $1.4\times10^{-8}$ & $2.2\times10^{-7}$\\
 3DMOT capture rate (atoms/s) & $3.5\times10^{8}$ & $5\times10^{10}$\\  [1ex]
 \hline
\end{tabular}
\caption{Optimised parameters for the 2D$^{+}$MOTs}
\label{tab:2DMOTKparameters}
\end{table}

\begin{table}[h!]
\centering
\setlength{\tabcolsep}{10pt} 
\renewcommand{\arraystretch}{1.5}

\begin{tabular}{c c c c} 
 \hline
 3DMOT parameters & $^{23}$Na & $^{39}$K \\ [0.5ex] 
 \hline\hline
 $\delta^{3D}_{cooling}$ ($\Gamma$)& -1.4 & -6.8 \\ 
 Total $I^{3D}_{cooling}$ ($I_{S}$) & 10 & 22.7 \\
 3DMOT field gradient (G/cm) & 17.6 & 18.5 \\
 $I^{3D}_{repumping}/I^{3D}_{cooling}$ & 0.225 & 0.75 \\ 
 Total number of atoms &  $5.8\times10^{8}$ &  $3\times10^{10}$ \\ [1ex]
 \hline
\end{tabular}
\caption{Optimised parameters for the 3DMOT}
\label{3D_table}
\end{table}


The dependence of the capture rate of the $^{39}$K-MOT and the $^{23}$Na-MOT on the cooling beam intensity of the corresponding 2D$^{+}$MOT is depicted in Fig.~\ref{figure:Fig5}. The curve almost linearly increases with the beam power without a clear indication of saturation. The increase is due to two effects: First, the 2D$^+$MOT capture velocity increases with laser power due to the power broadening of the atomic spectral lines. Second, the scattering force increases, resulting in steeper transverse confinement, which facilitates the injection of the atoms into the differential pumping tube. The absence of saturation demonstrates that light-induced collisions for the used range of laser powers are negligible. As the rate for light-induced collisions depends on the atom number density in the 2D$^+$MOT, the absence of saturation effect implies that the atomic density in the 2D$^+$MOT is low due to the absence of three-dimensional confinement. This qualitative description given above is supported well using our numerical simulation results which agree well with the experimental observation as evident in Fig.~\ref{figure:Fig5}. The capture rate of atoms in the 3DMOTs is limited by the available laser power in our experiment.

Fig.~\ref{figure:Fig6} shows the dependence of the capture rate of the $^{39}$K-MOT and the $^{23}$Na-MOT on the intensity ratio between the cooling and repumping beams of the corresponding 2D$^{+}$MOT. The graph shows that the $^{39}$K-MOT and $^{23}$Na-MOT capture rate increases with increasing repumping intensity and that it saturates at high intensities of repumping beams. The dependence of the capture rate on the repumping beam intensity can be attributed to the branching ratio of the transition probabilities for the corresponding atomic transitions in $^{39}$K and $^{23}$Na atoms. In the case of $^{39}$K atoms, the hyperfine splitting of the excited states (dipole allowed from the $|F=2\rangle$ hyperfine ground state) is small (37.2 MHz) thereby increasing the probability of optical pumping to the lower hyperfine state $|F=1\rangle$ removing the atoms from the cooling transition. Hence, a relatively large intensity ratio between the repumping and cooling beams is required for the $^{39}$K atoms as compared to $^{23}$Na atoms where the same excited state hyperfine splitting is relatively large (92.6 MHz). 


The dependence of the capture rate of the $^{39}$K-MOT and the $^{23}$Na-MOT on the intensity ratio between the pushing and retarding beam of the corresponding 2D$^{+}$MOT is depicted in Fig.~\ref{figure:Fig7}. The curve has a maximum at $I_\textrm{\scriptsize push}/I_\textrm{\scriptsize retard}=8.1$ for $^{39}$K-MOT and 3.6 for $^{23}$Na-MOT.  The difference observed between the optimised ratio of the pushing and retarding beams for $^{39}$K and $^{23}$Na 2D$^{+}$MOTs can be attributed to the following argument. The $^{23}$Na atoms tend to diffuse out of the cooling volume due to lower mass and larger scattering rates compared to the $^{39}$K atoms. Therefore, we need to also efficiently cool the $^{23}$Na atoms in the longitudinal direction so that they spend a longer duration in the transverse cooling region. This is achieved by a smaller $I_\textrm{\scriptsize push}/I_\textrm{\scriptsize retard}$ ratio as compared to the $^{39}$K atoms.

The dependence of the capture rate of $^{39}$K-MOT and the $^{23}$Na-MOT on the magnetic field gradient of the corresponding 2D$^{+}$MOT is shown in  Fig.~\ref{figure:Fig8}. The curves show a maximum capture rate of the 3DMOT at the 2D$^{+}$MOT magnetic field gradient of 8 G/cm and 25 G/cm for $^{39}$K and $^{23}$Na atoms respectively depicting the optimized 2D$^{+}$MOT magnetic field gradient for efficient magneto-optical confinement in the transverse direction so as to pass through the differential pumping hole with minimum loss of atoms. We have compared this observation with the numerical simulation results and found that at low magnetic field gradients the agreement is good for both species. However, at high field the simple two-level model is no longer a valid approximation because the excited state manifold (spacing between $|F' = 1\rangle$ and $|F' = 3\rangle$ states) in case of Potassium is pretty small ($\sim$ 31 MHz) while for Sodium it is around $\sim$ 93 MHz. Therefore, the force on an atom away from the 2D-MOT center axis is significantly high for the high gradient case compared to what is calculated using a simple two-level model. This effect, in turn, increases the atomic beam divergence limiting the capture rate into the 3DMOT. Expectedly, this effect is more serious in case of Potassium than Sodium as is evident from the Fig.~\ref{figure:Fig8}.

Fig.~\ref{figure:Fig9} shows the dependence of the $^{39}$K-MOT and $^{23}$Na-MOT capture rate on the vapour pressure of $^{39}$K as well as $^{23}$Na atoms in the 2D$^+$MOT cell. The vapour pressure was measured by recording the absorption profile of a low-intensity probe, fitting it to Beer's law and applying the ideal gas equation. For the fit, the isotopic abundances in the cell needed to be taken into account as the small hyperfine splittings and isotopic shifts of the potassium atoms lead to a single Doppler absorption profile induced by the different transitions. As evident from the plots in Fig.~\ref{figure:Fig9}, the capture rate in the 3DMOT increases linearly with vapour pressure in the 2D$^+$MOT glass cell for both $^{39}$K and $^{23}$Na atoms for low partial vapour pressures. We observe an indication of saturation in the capture rate for $^{39}$K source at around $2\times10^{-7}$ mbar vapour pressure. The capture rate for $^{23}$Na 3DMOT is only limited by the available partial vapour pressure of $^{23}$Na atoms in the 2D$^+$MOT glass cell. The  $^{23}$Na partial vapour pressure was not measured using the weak probe absorption technique, rather it was inferred from the temperature of the air around the glass cell. 
\par
The optimised experimental parameters for the $^{39}$K and $^{23}$Na 2D$^+$MOTs  are summarised in Table~\ref{tab:2DMOTKparameters} and the optimised parameters for the $^{39}$K and $^{23}$Na 3D MOTs are summarised in Table~\ref{3D_table}.
\par
In order to vary the vapour pressure in the 2D$^{+}$MOT side for $^{39}$K atoms, the oven temperature was varied between 50$^o$C to 130$^o$C. The metal parts near the 2D$^{+}$MOT glass cell were kept mildly heated to 40$^o$C so as to prevent coating of  $^{39}$K atoms there and to facilitate $^{39}$K atomic vapour coming into the glass cell. In the case of $^{23}$Na 2D$^+$MOT, the oven was heated to 300$^o$C and the metal parts near the glass cell were heated to 100$^o$C. The region around the $^{23}$Na 2D$^+$MOT glass cell was kept heat-insulated using a heat-insulation blanket and two layers of aluminium foil while keeping sufficient openings for the cooling laser beams. Heating rods were inserted inside the insulated region to keep the air temperature around the 2D$^{+}$MOT glass cell between 60-75$^o$C. In the case of Na, this special arrangement is done (in contrast to $^{39}$K) because $^{23}$Na atoms have a particular tendency to stick to glass surfaces and the melting point of $^{23}$Na is also relatively high (98$^o$C). We have avoided heating the glass cell to a higher temperature so as to prevent degrading the vacuum at the 2D$^{+}$MOT side.

The total number of atoms captured in the $^{39}$K-MOT and $^{23}$Na-MOT as a function of time demonstrating the effect of switching on the LIAD (Light-induced atomic desorption) for the corresponding 2D$^{+}$ MOT is shown in Fig.~\ref{figure:Fig10}. In our experiment, the LIAD plays a significant role in increasing the vapour pressure of  $^{39}$K and $^{23}$Na atoms in their respective 2D$^+$MOT glass cells increasing the atomic beam flux and thereby the capture rate of atoms in the corresponding 3DMOT. We use commercially available high-power UV Light-Emitting diodes (LED) (centre wavelength 395 nm) (Thorlabs M395L5) for the $^{39}$K side and a 100 low-power UV LED array for the $^{23}$Na side. Both $^{39}$K and $^{23}$Na atoms are efficiently desorbed from the glass surface on shining the UV light thereby increasing the partial pressure of the atoms without affecting the overall vacuum in the 2D$^{+}$ MOT glass cells. This results in a significant increase in the 2D$^{+}$ MOT flux as well as the 3DMOT capture rate as shown in Fig.~\ref{figure:Fig10}. This is particularly useful because apart from the improved performance of the cold atom sources, the UV light also prevents the glass surface to be coated by $^{39}$K and $^{23}$Na atoms.

%




\section{Light-assisted interspecies cold collisions} \label{sec:Beta}
\begin{figure}[ht!]
\centering
\includegraphics[width=8.8cm]{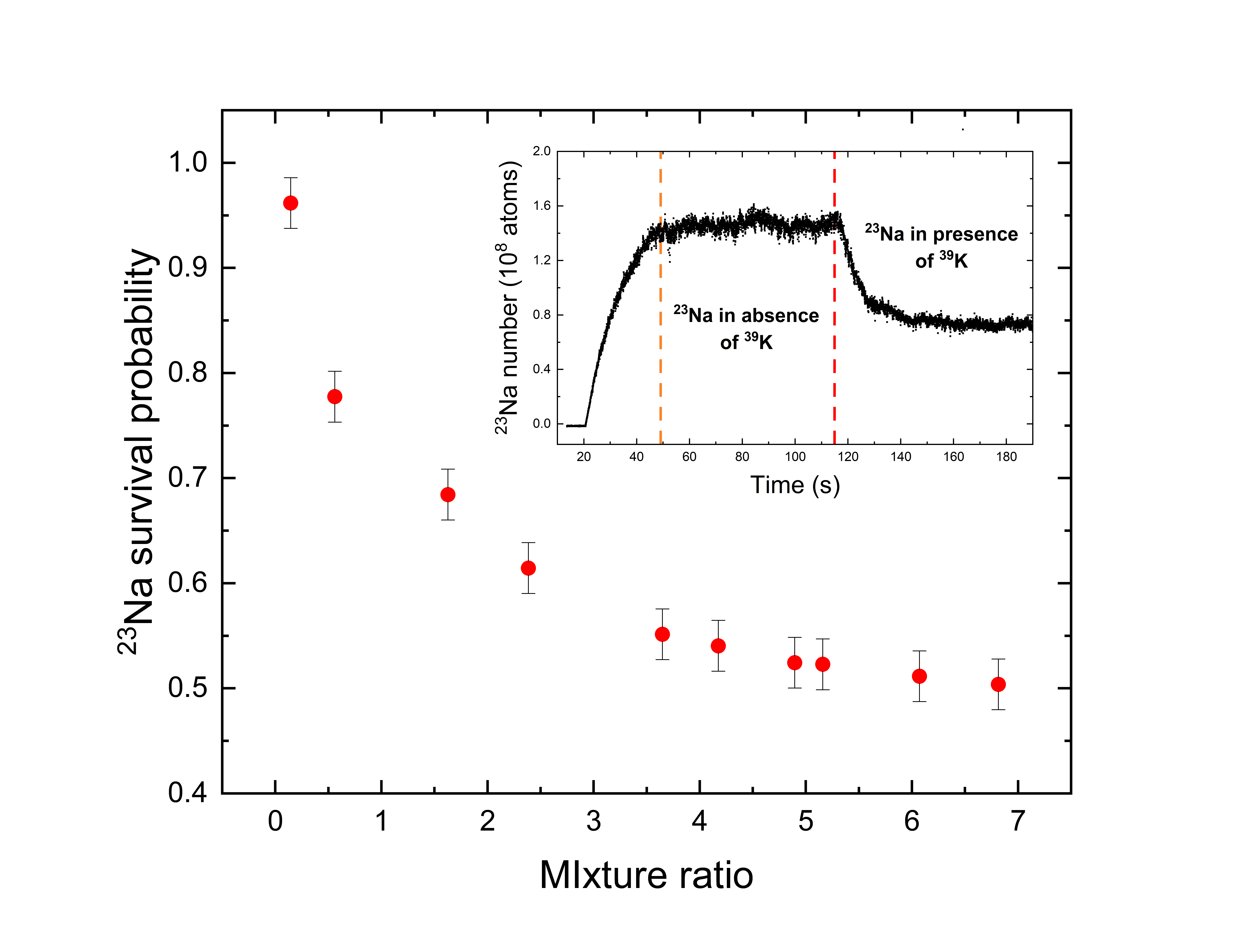}
\caption{The survival probability of $^{23}$Na atoms as a function of the ``Mixture ratio" (ratio between the number of $^{39}$K and $^{23}$Na atoms). Inset: A typical data of $^{23}$Na atom number in the 3DMOT in absence and presence of $^{39}$K 3DMOT loading.}
\label{figure:MixRatio}
\end{figure}

Finally, we report on the effect of cold collisions between $^{23}$Na and $^{39}$K atoms while they are simultaneously captured in the 3DMOT. The effects are considerable and may cause significant atom loss from the trap as evident from a typical experimental data shown in the inset of Fig.\ref{figure:MixRatio}, where we monitored only the number of $^{23}$Na trapped atoms while loading the $^{39}$K 3DMOT which overlaps with the $^{23}$Na 3DMOT in space. From a series of such data recorded with different $^{39}$K loading rate we experimentally find an interspecies loss-coefficient ($\beta_{NaK}$). We present the results as a survival probability of one species (reported for $^{23}$Na) in presence of the other species by calculating the total trap loss in the asymptotic limit (Fig.\ref{figure:MixRatio}). In this context, the survival probability is defined as the fraction of atoms remaining after interspecies light assisted collision is turned on. 

\par
It is worth noting that we observe as much as nearly 50$\%$ loss of $^{23}$Na atoms due to interspecies light assisted cold collisions. As the $^{39}$K numbers are increased in the trap by increasing the $^{39}$K 2D$^+$MOT flux, the interspecies collisions results in further loss of $^{23}$Na atoms which saturates above a mixture ratio (the ratio of the number of $^{39}$K atoms to the $^{23}$Na atoms in the trap) beyond 5. In this high Mixture ratio $^{39}$K can be considered a bath in which $^{23}$Na atoms move as ``impurities". As the bath size is increased we reach a constant density limit for $^{39}$K cloud, and the interspecies collision also reaches a steady state value. For typical dual species overlapping 3DMOTs of $^{23}$Na and $^{39}$K, we find the interspecies loss co-efficient, $\beta_{NaK} = 2 \times 10^{-12}$ cm$^3$/sec (using the semi-classical approach described in \cite{2001_Telles}). In comparison, the intra-species light-assisted collision rates for $^{23}$Na - $^{23}$Na is, $\beta_{NaNa}$ = 2.1 $\times 10^{-11}$ cm$^3$/sec and for $^{39}$K - $^{39}$K is, $\beta_{KK}$ = 3 $\times 10^{-11}$ cm$^3$/sec. These measurements are performed using the typical single species loading curves in the 3DMOT using a similar measurement technique described in  \cite{1999_Arimondo}.  
\par
\begin{figure}[ht]
\centering
\includegraphics[width=8.8cm]{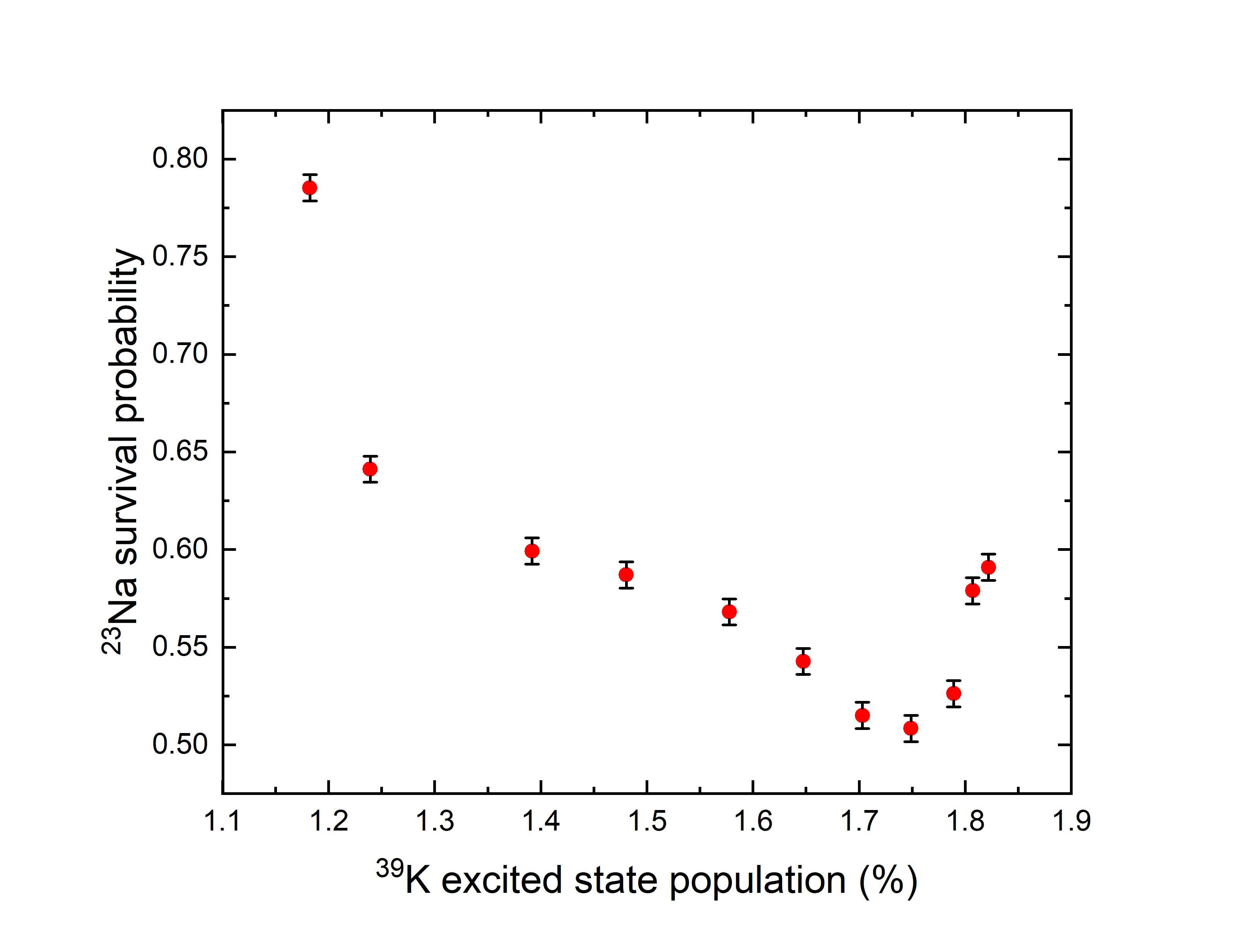}
\caption{The survival probability of $^{23}$Na atoms as a function of the excited state population of the $^{39}$K atoms.}
\label{figure:ESpop}
\end{figure}

 
In order to verify the above argument physically, we varied the interspecies loss-coefficient ($\beta_{NaK}$) by varying the excited state population of the bath atoms ($^{39}$K). Experimentally, this is done by controlling the repump laser power in 3DMOT which regulates the population of the $^{39}$K atoms in the $|F=2\rangle$ state. The resulting data is presented in Fig.\ref{figure:ESpop}. The observation indicates that as the excited state population is increased, the interspecies collision rate also increases resulting in further decrease in the number of minority species ($^{23}$Na) atoms. Interestingly, beyond a certain excited state fraction of around 1.75$\%$ in $^{39}$K atoms the survival probability of the $^{23}$Na atoms actually increases again. This is because, at such high $^{39}$K excited state population, $^{39}$K-$^{39}$K collisions (governed through $C_3/R^3$ potential) dominates over the $^{23}$Na-$^{39}$K collisions (which is governed through $C_6/R^6$ potential)\cite{1999_Santos}. Here, $R$ is the inter-atomic separation and $C_3$ and $C_6$ are the resonant and off-resonant dipole-dipole interaction co-efficients, respectively.  This effect results in the decrease of the density of the majority species ($^{39}$K) and hence the interspecies collision rates. This observation suggests that the interspecies collision can be tuned and significantly reduced by controlling the excited state population of the bath. This way, we could vary the $\beta_{NaK}$ in a range between 1.7 - 3.3 $\times 10^{-12}$ cm$^3$/sec.

\section{Conclusion and Outlook} \label{sec:Conclusion_Outlook}
We report the design, implementation and characterisation of two 2D$^+$MOT sources of cold atomic beam for $^{39}$K and $^{23}$Na atoms.  We characterize the performance of the two 2D$^+$MOT via measurements of the capture rate of atoms in the corresponding 3DMOT. We studied both experimentally as well as using a comprehensive numerical simulation, the dependence of the capture rate of cold atoms in the 3DMOT on the various 2D$^+$MOT parameters for both $^{39}$K and $^{23}$Na atoms. In this way, we obtain the optimized values of these parameters to ensure the best performance of the cold atomic beam sources. Joint performance of the overlapping $^{23}$Na and $^{39}$K MOTs have been studied and interspecies light-assisted collisional processes are reported along with the interspecies collision rates. 
\par
The cold atomic mixture will be cooled further via sub-Doppler cooling and loaded into a magnetic trap. Subsequently, the atoms would be transported to a `science cell' located at a distance of about 51 cm from the MOT centre using a magnetic transport and thereafter loaded into a QUIC magnetic trap for RF evaporative cooling to reach simultaneous quantum degeneracy of $^{23}$Na and $^{39}$K atoms. The quantum degenerate mixture would thereafter be transferred into optical traps and optical lattices \cite{2008_CS} for further experiments on quantum simulation of interacting many-body systems \cite{Bloch2012}. The response function of the cold atoms \cite{2022_Response} can be studied in the presence of inter-species interaction exploring a rich parameter space for the measurements. The spin-exchange interaction between hetero-nuclear cold atomic mixtures can be studied using Faraday rotation fluctuation measurements \cite{2021_Swar}. Our compact and versatile setup for realizing cold atomic beam sources of $^{39}$K and $^{23}$Na atoms will be useful for experiments on a quantum degenerate mixture of $^{39}$K and $^{23}$Na atoms as well as can be utilised for a variety of quantum technology experiments such as quantum metrology \cite{2016_Lewenstein}, quantum simulation \cite{2014_Nori} and quantum sensing \cite{2018_Pelegri}.  

\vspace{5mm}

\begin{acknowledgments}
This work was partially supported by the Ministry of Electronics and Information Technology (MeitY), Government of India, through the Center for Excellence in Quantum Technology, under Grant4(7)/2020-ITEA. S. R acknowledges funding from the Department of Science and Technology, India, via the WOS-A project grant no. SR/WOS-A/PM-59/2019. We acknowledge Meena M. S., Md. Ibrahim, S. Barui, Sujatha S, S. Bhar, B. B. Boruah, S. Bagchi, the RRI mechanical workshop, RRI HPC facility,  for the instruments and assistance with the experiments.

\end{acknowledgments}

\section*{Data Availability Statement}
The data that support the findings of this study are available from the corresponding author upon reasonable request.

\bibliography{doubleref}

\begin{thebibliography}{69}%
\makeatletter
\providecommand \@ifxundefined [1]{%
 \@ifx{#1\undefined}
}%
\providecommand \@ifnum [1]{%
 \ifnum #1\expandafter \@firstoftwo
 \else \expandafter \@secondoftwo
 \fi
}%
\providecommand \@ifx [1]{%
 \ifx #1\expandafter \@firstoftwo
 \else \expandafter \@secondoftwo
 \fi
}%
\providecommand \natexlab [1]{#1}%
\providecommand \enquote  [1]{``#1''}%
\providecommand \bibnamefont  [1]{#1}%
\providecommand \bibfnamefont [1]{#1}%
\providecommand \citenamefont [1]{#1}%
\providecommand \href@noop [0]{\@secondoftwo}%
\providecommand \href [0]{\begingroup \@sanitize@url \@href}%
\providecommand \@href[1]{\@@startlink{#1}\@@href}%
\providecommand \@@href[1]{\endgroup#1\@@endlink}%
\providecommand \@sanitize@url [0]{\catcode `\\12\catcode `\$12\catcode
  `\&12\catcode `\#12\catcode `\^12\catcode `\_12\catcode `\%12\relax}%
\providecommand \@@startlink[1]{}%
\providecommand \@@endlink[0]{}%
\providecommand \url  [0]{\begingroup\@sanitize@url \@url }%
\providecommand \@url [1]{\endgroup\@href {#1}{\urlprefix }}%
\providecommand \urlprefix  [0]{URL }%
\providecommand \Eprint [0]{\href }%
\providecommand \doibase [0]{http://dx.doi.org/}%
\providecommand \selectlanguage [0]{\@gobble}%
\providecommand \bibinfo  [0]{\@secondoftwo}%
\providecommand \bibfield  [0]{\@secondoftwo}%
\providecommand \translation [1]{[#1]}%
\providecommand \BibitemOpen [0]{}%
\providecommand \bibitemStop [0]{}%
\providecommand \bibitemNoStop [0]{.\EOS\space}%
\providecommand \EOS [0]{\spacefactor3000\relax}%
\providecommand \BibitemShut  [1]{\csname bibitem#1\endcsname}%
\let\auto@bib@innerbib\@empty
\bibitem [{\citenamefont {Bloch}, \citenamefont {Dalibard},\ and\ \citenamefont
  {Zwerger}(2008)}]{BlochRMP2008}%
  \BibitemOpen
  \bibfield  {author} {\bibinfo {author} {\bibfnamefont {I.}~\bibnamefont
  {Bloch}}, \bibinfo {author} {\bibfnamefont {J.}~\bibnamefont {Dalibard}}, \
  and\ \bibinfo {author} {\bibfnamefont {W.}~\bibnamefont {Zwerger}},\
  }\bibfield  {title} {\enquote {\bibinfo {title} {Many-body physics with
  ultracold gases},}\ }\href {\doibase 10.1103/RevModPhys.80.885} {\bibfield
  {journal} {\bibinfo  {journal} {Rev. Mod. Phys.}\ }\textbf {\bibinfo {volume}
  {80}},\ \bibinfo {pages} {885--964} (\bibinfo {year} {2008})}\BibitemShut
  {NoStop}%
\bibitem [{\citenamefont {Bloch}, \citenamefont {Dalibard},\ and\ \citenamefont
  {Nascimb{\`e}ne}(2012)}]{Bloch2012}%
  \BibitemOpen
  \bibfield  {author} {\bibinfo {author} {\bibfnamefont {I.}~\bibnamefont
  {Bloch}}, \bibinfo {author} {\bibfnamefont {J.}~\bibnamefont {Dalibard}}, \
  and\ \bibinfo {author} {\bibfnamefont {S.}~\bibnamefont {Nascimb{\`e}ne}},\
  }\bibfield  {title} {\enquote {\bibinfo {title} {Quantum simulations with
  ultracold quantum gases},}\ }\href {\doibase 10.1038/nphys2259} {\bibfield
  {journal} {\bibinfo  {journal} {Nature Physics}\ }\textbf {\bibinfo {volume}
  {8}},\ \bibinfo {pages} {267--276} (\bibinfo {year} {2012})}\BibitemShut
  {NoStop}%
\bibitem [{\citenamefont {Gross}\ and\ \citenamefont
  {Bloch}(2017)}]{Gross2017Science}%
  \BibitemOpen
  \bibfield  {author} {\bibinfo {author} {\bibfnamefont {C.}~\bibnamefont
  {Gross}}\ and\ \bibinfo {author} {\bibfnamefont {I.}~\bibnamefont {Bloch}},\
  }\bibfield  {title} {\enquote {\bibinfo {title} {Quantum simulations with
  ultracold atoms in optical lattices},}\ }\href {\doibase
  10.1126/science.aal3837} {\bibfield  {journal} {\bibinfo  {journal}
  {Science}\ }\textbf {\bibinfo {volume} {357}},\ \bibinfo {pages} {995--1001}
  (\bibinfo {year} {2017})},\ \Eprint
  {http://arxiv.org/abs/https://www.science.org/doi/pdf/10.1126/science.aal3837}
  {https://www.science.org/doi/pdf/10.1126/science.aal3837} \BibitemShut
  {NoStop}%
\bibitem [{\citenamefont {Sch{\"a}fer}\ \emph {et~al.}(2020)\citenamefont
  {Sch{\"a}fer}, \citenamefont {Fukuhara}, \citenamefont {Sugawa},
  \citenamefont {Takasu},\ and\ \citenamefont {Takahashi}}]{Schafer2020}%
  \BibitemOpen
  \bibfield  {author} {\bibinfo {author} {\bibfnamefont {F.}~\bibnamefont
  {Sch{\"a}fer}}, \bibinfo {author} {\bibfnamefont {T.}~\bibnamefont
  {Fukuhara}}, \bibinfo {author} {\bibfnamefont {S.}~\bibnamefont {Sugawa}},
  \bibinfo {author} {\bibfnamefont {Y.}~\bibnamefont {Takasu}}, \ and\ \bibinfo
  {author} {\bibfnamefont {Y.}~\bibnamefont {Takahashi}},\ }\bibfield  {title}
  {\enquote {\bibinfo {title} {Tools for quantum simulation with ultracold
  atoms in optical lattices},}\ }\href {\doibase 10.1038/s42254-020-0195-3}
  {\bibfield  {journal} {\bibinfo  {journal} {Nature Reviews Physics}\ }\textbf
  {\bibinfo {volume} {2}},\ \bibinfo {pages} {411--425} (\bibinfo {year}
  {2020})}\BibitemShut {NoStop}%
\bibitem [{\citenamefont {Denschlag}\ \emph {et~al.}(2002)\citenamefont
  {Denschlag}, \citenamefont {Simsarian}, \citenamefont {H{\"a}ffner},
  \citenamefont {McKenzie}, \citenamefont {Browaeys}, \citenamefont {Cho},
  \citenamefont {Helmerson}, \citenamefont {Rolston},\ and\ \citenamefont
  {Phillips}}]{Denschlag_2002}%
  \BibitemOpen
  \bibfield  {author} {\bibinfo {author} {\bibfnamefont {J.~H.}\ \bibnamefont
  {Denschlag}}, \bibinfo {author} {\bibfnamefont {J.~E.}\ \bibnamefont
  {Simsarian}}, \bibinfo {author} {\bibfnamefont {H.}~\bibnamefont
  {H{\"a}ffner}}, \bibinfo {author} {\bibfnamefont {C.}~\bibnamefont
  {McKenzie}}, \bibinfo {author} {\bibfnamefont {A.}~\bibnamefont {Browaeys}},
  \bibinfo {author} {\bibfnamefont {D.}~\bibnamefont {Cho}}, \bibinfo {author}
  {\bibfnamefont {K.}~\bibnamefont {Helmerson}}, \bibinfo {author}
  {\bibfnamefont {S.~L.}\ \bibnamefont {Rolston}}, \ and\ \bibinfo {author}
  {\bibfnamefont {W.~D.}\ \bibnamefont {Phillips}},\ }\bibfield  {title}
  {\enquote {\bibinfo {title} {A bose-einstein condensate in an optical
  lattice},}\ }\href {\doibase 10.1088/0953-4075/35/14/307} {\bibfield
  {journal} {\bibinfo  {journal} {Journal of Physics B: Atomic, Molecular and
  Optical Physics}\ }\textbf {\bibinfo {volume} {35}},\ \bibinfo {pages}
  {3095--3110} (\bibinfo {year} {2002})}\BibitemShut {NoStop}%
\bibitem [{\citenamefont {Ye}\ \emph {et~al.}(2007)\citenamefont {Ye},
  \citenamefont {Blatt}, \citenamefont {Boyd}, \citenamefont {Foreman},
  \citenamefont {Hudson}, \citenamefont {Ido}, \citenamefont {Lev},
  \citenamefont {Ludlow}, \citenamefont {Sawyer}, \citenamefont {Stuhl},\ and\
  \citenamefont {Zelinsky}}]{YeIJMPD2007}%
  \BibitemOpen
  \bibfield  {author} {\bibinfo {author} {\bibfnamefont {J.}~\bibnamefont
  {Ye}}, \bibinfo {author} {\bibfnamefont {S.}~\bibnamefont {Blatt}}, \bibinfo
  {author} {\bibfnamefont {M.~M.}\ \bibnamefont {Boyd}}, \bibinfo {author}
  {\bibfnamefont {S.~M.}\ \bibnamefont {Foreman}}, \bibinfo {author}
  {\bibfnamefont {E.~R.}\ \bibnamefont {Hudson}}, \bibinfo {author}
  {\bibfnamefont {T.}~\bibnamefont {Ido}}, \bibinfo {author} {\bibfnamefont
  {B.}~\bibnamefont {Lev}}, \bibinfo {author} {\bibfnamefont {A.~D.}\
  \bibnamefont {Ludlow}}, \bibinfo {author} {\bibfnamefont {B.~C.}\
  \bibnamefont {Sawyer}}, \bibinfo {author} {\bibfnamefont {B.}~\bibnamefont
  {Stuhl}}, \ and\ \bibinfo {author} {\bibfnamefont {T.}~\bibnamefont
  {Zelinsky}},\ }\bibfield  {title} {\enquote {\bibinfo {title} {Precision
  measurement and frequency metrology with ultracold atoms measurement based on
  ultracold atoms and cold molecules},}\ }\href {\doibase
  10.1142/S0218271807011826} {\bibfield  {journal} {\bibinfo  {journal}
  {International Journal of Modern Physics D}\ }\textbf {\bibinfo {volume}
  {16}},\ \bibinfo {pages} {2481--2494} (\bibinfo {year} {2007})},\ \Eprint
  {http://arxiv.org/abs/https://doi.org/10.1142/S0218271807011826}
  {https://doi.org/10.1142/S0218271807011826} \BibitemShut {NoStop}%
\bibitem [{\citenamefont {Zhang}\ and\ \citenamefont {Ye}(2016)}]{Zhang2016}%
  \BibitemOpen
  \bibfield  {author} {\bibinfo {author} {\bibfnamefont {X.}~\bibnamefont
  {Zhang}}\ and\ \bibinfo {author} {\bibfnamefont {J.}~\bibnamefont {Ye}},\
  }\bibfield  {title} {\enquote {\bibinfo {title} {Precision measurement and
  frequency metrology with ultracold atoms},}\ }\href {\doibase
  10.1093/nsr/nww013} {\bibfield  {journal} {\bibinfo  {journal} {National
  Science Review}\ }\textbf {\bibinfo {volume} {3}},\ \bibinfo {pages} {nww013}
  (\bibinfo {year} {2016})}\BibitemShut {NoStop}%
\bibitem [{\citenamefont {Myatt}\ \emph {et~al.}(1997)\citenamefont {Myatt},
  \citenamefont {Burt}, \citenamefont {Ghrist}, \citenamefont {Cornell},\ and\
  \citenamefont {Wieman}}]{MyattPRL1997}%
  \BibitemOpen
  \bibfield  {author} {\bibinfo {author} {\bibfnamefont {C.~J.}\ \bibnamefont
  {Myatt}}, \bibinfo {author} {\bibfnamefont {E.~A.}\ \bibnamefont {Burt}},
  \bibinfo {author} {\bibfnamefont {R.~W.}\ \bibnamefont {Ghrist}}, \bibinfo
  {author} {\bibfnamefont {E.~A.}\ \bibnamefont {Cornell}}, \ and\ \bibinfo
  {author} {\bibfnamefont {C.~E.}\ \bibnamefont {Wieman}},\ }\bibfield  {title}
  {\enquote {\bibinfo {title} {Production of two overlapping bose-einstein
  condensates by sympathetic cooling},}\ }\href {\doibase
  10.1103/PhysRevLett.78.586} {\bibfield  {journal} {\bibinfo  {journal} {Phys.
  Rev. Lett.}\ }\textbf {\bibinfo {volume} {78}},\ \bibinfo {pages} {586--589}
  (\bibinfo {year} {1997})}\BibitemShut {NoStop}%
\bibitem [{\citenamefont {Hall}\ \emph {et~al.}(1998)\citenamefont {Hall},
  \citenamefont {Matthews}, \citenamefont {Ensher}, \citenamefont {Wieman},\
  and\ \citenamefont {Cornell}}]{HallPRL1998}%
  \BibitemOpen
  \bibfield  {author} {\bibinfo {author} {\bibfnamefont {D.~S.}\ \bibnamefont
  {Hall}}, \bibinfo {author} {\bibfnamefont {M.~R.}\ \bibnamefont {Matthews}},
  \bibinfo {author} {\bibfnamefont {J.~R.}\ \bibnamefont {Ensher}}, \bibinfo
  {author} {\bibfnamefont {C.~E.}\ \bibnamefont {Wieman}}, \ and\ \bibinfo
  {author} {\bibfnamefont {E.~A.}\ \bibnamefont {Cornell}},\ }\bibfield
  {title} {\enquote {\bibinfo {title} {Dynamics of component separation in a
  binary mixture of bose-einstein condensates},}\ }\href {\doibase
  10.1103/PhysRevLett.81.1539} {\bibfield  {journal} {\bibinfo  {journal}
  {Phys. Rev. Lett.}\ }\textbf {\bibinfo {volume} {81}},\ \bibinfo {pages}
  {1539--1542} (\bibinfo {year} {1998})}\BibitemShut {NoStop}%
\bibitem [{\citenamefont {Stenger}\ \emph {et~al.}(1998)\citenamefont
  {Stenger}, \citenamefont {Inouye}, \citenamefont {Stamper-Kurn},
  \citenamefont {Miesner}, \citenamefont {Chikkatur},\ and\ \citenamefont
  {Ketterle}}]{Stenger_1998}%
  \BibitemOpen
  \bibfield  {author} {\bibinfo {author} {\bibfnamefont {J.}~\bibnamefont
  {Stenger}}, \bibinfo {author} {\bibfnamefont {S.}~\bibnamefont {Inouye}},
  \bibinfo {author} {\bibfnamefont {D.~M.}\ \bibnamefont {Stamper-Kurn}},
  \bibinfo {author} {\bibfnamefont {H.-J.}\ \bibnamefont {Miesner}}, \bibinfo
  {author} {\bibfnamefont {A.~P.}\ \bibnamefont {Chikkatur}}, \ and\ \bibinfo
  {author} {\bibfnamefont {W.}~\bibnamefont {Ketterle}},\ }\bibfield  {title}
  {\enquote {\bibinfo {title} {Spin domains in ground-state bose--einstein
  condensates},}\ }\href {\doibase 10.1038/24567} {\bibfield  {journal}
  {\bibinfo  {journal} {Nature}\ }\textbf {\bibinfo {volume} {396}},\ \bibinfo
  {pages} {345--348} (\bibinfo {year} {1998})}\BibitemShut {NoStop}%
\bibitem [{\citenamefont {Modugno}\ \emph {et~al.}(2001)\citenamefont
  {Modugno}, \citenamefont {Ferrari}, \citenamefont {Roati}, \citenamefont
  {Brecha}, \citenamefont {Simoni},\ and\ \citenamefont
  {Inguscio}}]{Modugno1320}%
  \BibitemOpen
  \bibfield  {author} {\bibinfo {author} {\bibfnamefont {G.}~\bibnamefont
  {Modugno}}, \bibinfo {author} {\bibfnamefont {G.}~\bibnamefont {Ferrari}},
  \bibinfo {author} {\bibfnamefont {G.}~\bibnamefont {Roati}}, \bibinfo
  {author} {\bibfnamefont {R.~J.}\ \bibnamefont {Brecha}}, \bibinfo {author}
  {\bibfnamefont {A.}~\bibnamefont {Simoni}}, \ and\ \bibinfo {author}
  {\bibfnamefont {M.}~\bibnamefont {Inguscio}},\ }\bibfield  {title} {\enquote
  {\bibinfo {title} {Bose-einstein condensation of potassium atoms by
  sympathetic cooling},}\ }\href {\doibase 10.1126/science.1066687} {\bibfield
  {journal} {\bibinfo  {journal} {Science}\ }\textbf {\bibinfo {volume}
  {294}},\ \bibinfo {pages} {1320--1322} (\bibinfo {year} {2001})},\ \Eprint
  {http://arxiv.org/abs/https://science.sciencemag.org/content/294/5545/1320.full.pdf}
  {https://science.sciencemag.org/content/294/5545/1320.full.pdf} \BibitemShut
  {NoStop}%
\bibitem [{\citenamefont {Hadzibabic}\ \emph {et~al.}(2002)\citenamefont
  {Hadzibabic}, \citenamefont {Stan}, \citenamefont {Dieckmann}, \citenamefont
  {Gupta}, \citenamefont {Zwierlein}, \citenamefont {G\"orlitz},\ and\
  \citenamefont {Ketterle}}]{HadzibabicPRL2002}%
  \BibitemOpen
  \bibfield  {author} {\bibinfo {author} {\bibfnamefont {Z.}~\bibnamefont
  {Hadzibabic}}, \bibinfo {author} {\bibfnamefont {C.~A.}\ \bibnamefont
  {Stan}}, \bibinfo {author} {\bibfnamefont {K.}~\bibnamefont {Dieckmann}},
  \bibinfo {author} {\bibfnamefont {S.}~\bibnamefont {Gupta}}, \bibinfo
  {author} {\bibfnamefont {M.~W.}\ \bibnamefont {Zwierlein}}, \bibinfo {author}
  {\bibfnamefont {A.}~\bibnamefont {G\"orlitz}}, \ and\ \bibinfo {author}
  {\bibfnamefont {W.}~\bibnamefont {Ketterle}},\ }\bibfield  {title} {\enquote
  {\bibinfo {title} {Two-species mixture of quantum degenerate bose and fermi
  gases},}\ }\href {\doibase 10.1103/PhysRevLett.88.160401} {\bibfield
  {journal} {\bibinfo  {journal} {Phys. Rev. Lett.}\ }\textbf {\bibinfo
  {volume} {88}},\ \bibinfo {pages} {160401} (\bibinfo {year}
  {2002})}\BibitemShut {NoStop}%
\bibitem [{\citenamefont {Modugno}\ \emph {et~al.}(2002)\citenamefont
  {Modugno}, \citenamefont {Modugno}, \citenamefont {Riboli}, \citenamefont
  {Roati},\ and\ \citenamefont {Inguscio}}]{ModugnoPRL2002}%
  \BibitemOpen
  \bibfield  {author} {\bibinfo {author} {\bibfnamefont {G.}~\bibnamefont
  {Modugno}}, \bibinfo {author} {\bibfnamefont {M.}~\bibnamefont {Modugno}},
  \bibinfo {author} {\bibfnamefont {F.}~\bibnamefont {Riboli}}, \bibinfo
  {author} {\bibfnamefont {G.}~\bibnamefont {Roati}}, \ and\ \bibinfo {author}
  {\bibfnamefont {M.}~\bibnamefont {Inguscio}},\ }\bibfield  {title} {\enquote
  {\bibinfo {title} {Two atomic species superfluid},}\ }\href {\doibase
  10.1103/PhysRevLett.89.190404} {\bibfield  {journal} {\bibinfo  {journal}
  {Phys. Rev. Lett.}\ }\textbf {\bibinfo {volume} {89}},\ \bibinfo {pages}
  {190404} (\bibinfo {year} {2002})}\BibitemShut {NoStop}%
\bibitem [{\citenamefont {Regal}, \citenamefont {Greiner},\ and\ \citenamefont
  {Jin}(2004)}]{RegalPRL2004}%
  \BibitemOpen
  \bibfield  {author} {\bibinfo {author} {\bibfnamefont {C.~A.}\ \bibnamefont
  {Regal}}, \bibinfo {author} {\bibfnamefont {M.}~\bibnamefont {Greiner}}, \
  and\ \bibinfo {author} {\bibfnamefont {D.~S.}\ \bibnamefont {Jin}},\
  }\bibfield  {title} {\enquote {\bibinfo {title} {Observation of resonance
  condensation of fermionic atom pairs},}\ }\href {\doibase
  10.1103/PhysRevLett.92.040403} {\bibfield  {journal} {\bibinfo  {journal}
  {Phys. Rev. Lett.}\ }\textbf {\bibinfo {volume} {92}},\ \bibinfo {pages}
  {040403} (\bibinfo {year} {2004})}\BibitemShut {NoStop}%
\bibitem [{\citenamefont {Silber}\ \emph {et~al.}(2005)\citenamefont {Silber},
  \citenamefont {G\"unther}, \citenamefont {Marzok}, \citenamefont {Deh},
  \citenamefont {Courteille},\ and\ \citenamefont
  {Zimmermann}}]{SilberPRL2005}%
  \BibitemOpen
  \bibfield  {author} {\bibinfo {author} {\bibfnamefont {C.}~\bibnamefont
  {Silber}}, \bibinfo {author} {\bibfnamefont {S.}~\bibnamefont {G\"unther}},
  \bibinfo {author} {\bibfnamefont {C.}~\bibnamefont {Marzok}}, \bibinfo
  {author} {\bibfnamefont {B.}~\bibnamefont {Deh}}, \bibinfo {author}
  {\bibfnamefont {P.~W.}\ \bibnamefont {Courteille}}, \ and\ \bibinfo {author}
  {\bibfnamefont {C.}~\bibnamefont {Zimmermann}},\ }\bibfield  {title}
  {\enquote {\bibinfo {title} {Quantum-degenerate mixture of fermionic lithium
  and bosonic rubidium gases},}\ }\href {\doibase
  10.1103/PhysRevLett.95.170408} {\bibfield  {journal} {\bibinfo  {journal}
  {Phys. Rev. Lett.}\ }\textbf {\bibinfo {volume} {95}},\ \bibinfo {pages}
  {170408} (\bibinfo {year} {2005})}\BibitemShut {NoStop}%
\bibitem [{\citenamefont {G\"unter}\ \emph {et~al.}(2006)\citenamefont
  {G\"unter}, \citenamefont {St\"oferle}, \citenamefont {Moritz}, \citenamefont
  {K\"ohl},\ and\ \citenamefont {Esslinger}}]{Gunter_PRL_2006}%
  \BibitemOpen
  \bibfield  {author} {\bibinfo {author} {\bibfnamefont {K.}~\bibnamefont
  {G\"unter}}, \bibinfo {author} {\bibfnamefont {T.}~\bibnamefont
  {St\"oferle}}, \bibinfo {author} {\bibfnamefont {H.}~\bibnamefont {Moritz}},
  \bibinfo {author} {\bibfnamefont {M.}~\bibnamefont {K\"ohl}}, \ and\ \bibinfo
  {author} {\bibfnamefont {T.}~\bibnamefont {Esslinger}},\ }\bibfield  {title}
  {\enquote {\bibinfo {title} {Bose-fermi mixtures in a three-dimensional
  optical lattice},}\ }\href {\doibase 10.1103/PhysRevLett.96.180402}
  {\bibfield  {journal} {\bibinfo  {journal} {Phys. Rev. Lett.}\ }\textbf
  {\bibinfo {volume} {96}},\ \bibinfo {pages} {180402} (\bibinfo {year}
  {2006})}\BibitemShut {NoStop}%
\bibitem [{\citenamefont {Papp}, \citenamefont {Pino},\ and\ \citenamefont
  {Wieman}(2008)}]{PappPRL2008}%
  \BibitemOpen
  \bibfield  {author} {\bibinfo {author} {\bibfnamefont {S.~B.}\ \bibnamefont
  {Papp}}, \bibinfo {author} {\bibfnamefont {J.~M.}\ \bibnamefont {Pino}}, \
  and\ \bibinfo {author} {\bibfnamefont {C.~E.}\ \bibnamefont {Wieman}},\
  }\bibfield  {title} {\enquote {\bibinfo {title} {Tunable miscibility in a
  dual-species bose-einstein condensate},}\ }\href {\doibase
  10.1103/PhysRevLett.101.040402} {\bibfield  {journal} {\bibinfo  {journal}
  {Phys. Rev. Lett.}\ }\textbf {\bibinfo {volume} {101}},\ \bibinfo {pages}
  {040402} (\bibinfo {year} {2008})}\BibitemShut {NoStop}%
\bibitem [{\citenamefont {Ospelkaus}\ \emph
  {et~al.}(2006{\natexlab{a}})\citenamefont {Ospelkaus}, \citenamefont
  {Ospelkaus}, \citenamefont {Wille}, \citenamefont {Succo}, \citenamefont
  {Ernst}, \citenamefont {Sengstock},\ and\ \citenamefont
  {Bongs}}]{Ospelkaus_PRL_2006}%
  \BibitemOpen
  \bibfield  {author} {\bibinfo {author} {\bibfnamefont {S.}~\bibnamefont
  {Ospelkaus}}, \bibinfo {author} {\bibfnamefont {C.}~\bibnamefont
  {Ospelkaus}}, \bibinfo {author} {\bibfnamefont {O.}~\bibnamefont {Wille}},
  \bibinfo {author} {\bibfnamefont {M.}~\bibnamefont {Succo}}, \bibinfo
  {author} {\bibfnamefont {P.}~\bibnamefont {Ernst}}, \bibinfo {author}
  {\bibfnamefont {K.}~\bibnamefont {Sengstock}}, \ and\ \bibinfo {author}
  {\bibfnamefont {K.}~\bibnamefont {Bongs}},\ }\bibfield  {title} {\enquote
  {\bibinfo {title} {Localization of bosonic atoms by fermionic impurities in a
  three-dimensional optical lattice},}\ }\href {\doibase
  10.1103/PhysRevLett.96.180403} {\bibfield  {journal} {\bibinfo  {journal}
  {Phys. Rev. Lett.}\ }\textbf {\bibinfo {volume} {96}},\ \bibinfo {pages}
  {180403} (\bibinfo {year} {2006}{\natexlab{a}})}\BibitemShut {NoStop}%
\bibitem [{\citenamefont {Catani}\ \emph {et~al.}(2008)\citenamefont {Catani},
  \citenamefont {De~Sarlo}, \citenamefont {Barontini}, \citenamefont
  {Minardi},\ and\ \citenamefont {Inguscio}}]{Catani_PRA_2008}%
  \BibitemOpen
  \bibfield  {author} {\bibinfo {author} {\bibfnamefont {J.}~\bibnamefont
  {Catani}}, \bibinfo {author} {\bibfnamefont {L.}~\bibnamefont {De~Sarlo}},
  \bibinfo {author} {\bibfnamefont {G.}~\bibnamefont {Barontini}}, \bibinfo
  {author} {\bibfnamefont {F.}~\bibnamefont {Minardi}}, \ and\ \bibinfo
  {author} {\bibfnamefont {M.}~\bibnamefont {Inguscio}},\ }\bibfield  {title}
  {\enquote {\bibinfo {title} {Degenerate bose-bose mixture in a
  three-dimensional optical lattice},}\ }\href {\doibase
  10.1103/PhysRevA.77.011603} {\bibfield  {journal} {\bibinfo  {journal} {Phys.
  Rev. A}\ }\textbf {\bibinfo {volume} {77}},\ \bibinfo {pages} {011603}
  (\bibinfo {year} {2008})}\BibitemShut {NoStop}%
\bibitem [{\citenamefont {Taglieber}\ \emph {et~al.}(2008)\citenamefont
  {Taglieber}, \citenamefont {Voigt}, \citenamefont {Aoki}, \citenamefont
  {H\"ansch},\ and\ \citenamefont {Dieckmann}}]{TaglieberPRL2008}%
  \BibitemOpen
  \bibfield  {author} {\bibinfo {author} {\bibfnamefont {M.}~\bibnamefont
  {Taglieber}}, \bibinfo {author} {\bibfnamefont {A.-C.}\ \bibnamefont
  {Voigt}}, \bibinfo {author} {\bibfnamefont {T.}~\bibnamefont {Aoki}},
  \bibinfo {author} {\bibfnamefont {T.~W.}\ \bibnamefont {H\"ansch}}, \ and\
  \bibinfo {author} {\bibfnamefont {K.}~\bibnamefont {Dieckmann}},\ }\bibfield
  {title} {\enquote {\bibinfo {title} {Quantum degenerate two-species
  fermi-fermi mixture coexisting with a bose-einstein condensate},}\ }\href
  {\doibase 10.1103/PhysRevLett.100.010401} {\bibfield  {journal} {\bibinfo
  {journal} {Phys. Rev. Lett.}\ }\textbf {\bibinfo {volume} {100}},\ \bibinfo
  {pages} {010401} (\bibinfo {year} {2008})}\BibitemShut {NoStop}%
\bibitem [{\citenamefont {McCarron}\ \emph {et~al.}(2011)\citenamefont
  {McCarron}, \citenamefont {Cho}, \citenamefont {Jenkin}, \citenamefont
  {K\"oppinger},\ and\ \citenamefont {Cornish}}]{MacCarronPRA2011}%
  \BibitemOpen
  \bibfield  {author} {\bibinfo {author} {\bibfnamefont {D.~J.}\ \bibnamefont
  {McCarron}}, \bibinfo {author} {\bibfnamefont {H.~W.}\ \bibnamefont {Cho}},
  \bibinfo {author} {\bibfnamefont {D.~L.}\ \bibnamefont {Jenkin}}, \bibinfo
  {author} {\bibfnamefont {M.~P.}\ \bibnamefont {K\"oppinger}}, \ and\ \bibinfo
  {author} {\bibfnamefont {S.~L.}\ \bibnamefont {Cornish}},\ }\bibfield
  {title} {\enquote {\bibinfo {title} {Dual-species bose-einstein condensate of
  $^{87}\mathrm{Rb}$ and $^{133}\mathrm{Cs}$},}\ }\href {\doibase
  10.1103/PhysRevA.84.011603} {\bibfield  {journal} {\bibinfo  {journal} {Phys.
  Rev. A}\ }\textbf {\bibinfo {volume} {84}},\ \bibinfo {pages} {011603}
  (\bibinfo {year} {2011})}\BibitemShut {NoStop}%
\bibitem [{\citenamefont {Pasquiou}\ \emph {et~al.}(2013)\citenamefont
  {Pasquiou}, \citenamefont {Bayerle}, \citenamefont {Tzanova}, \citenamefont
  {Stellmer}, \citenamefont {Szczepkowski}, \citenamefont {Parigger},
  \citenamefont {Grimm},\ and\ \citenamefont {Schreck}}]{PasquiouPRA2013}%
  \BibitemOpen
  \bibfield  {author} {\bibinfo {author} {\bibfnamefont {B.}~\bibnamefont
  {Pasquiou}}, \bibinfo {author} {\bibfnamefont {A.}~\bibnamefont {Bayerle}},
  \bibinfo {author} {\bibfnamefont {S.~M.}\ \bibnamefont {Tzanova}}, \bibinfo
  {author} {\bibfnamefont {S.}~\bibnamefont {Stellmer}}, \bibinfo {author}
  {\bibfnamefont {J.}~\bibnamefont {Szczepkowski}}, \bibinfo {author}
  {\bibfnamefont {M.}~\bibnamefont {Parigger}}, \bibinfo {author}
  {\bibfnamefont {R.}~\bibnamefont {Grimm}}, \ and\ \bibinfo {author}
  {\bibfnamefont {F.}~\bibnamefont {Schreck}},\ }\bibfield  {title} {\enquote
  {\bibinfo {title} {Quantum degenerate mixtures of strontium and rubidium
  atoms},}\ }\href {\doibase 10.1103/PhysRevA.88.023601} {\bibfield  {journal}
  {\bibinfo  {journal} {Phys. Rev. A}\ }\textbf {\bibinfo {volume} {88}},\
  \bibinfo {pages} {023601} (\bibinfo {year} {2013})}\BibitemShut {NoStop}%
\bibitem [{\citenamefont {Ferrier-Barbut}\ \emph {et~al.}(2014)\citenamefont
  {Ferrier-Barbut}, \citenamefont {Delehaye}, \citenamefont {Laurent},
  \citenamefont {Grier}, \citenamefont {Pierce}, \citenamefont {Rem},
  \citenamefont {Chevy},\ and\ \citenamefont {Salomon}}]{Barbutscience2014}%
  \BibitemOpen
  \bibfield  {author} {\bibinfo {author} {\bibfnamefont {I.}~\bibnamefont
  {Ferrier-Barbut}}, \bibinfo {author} {\bibfnamefont {M.}~\bibnamefont
  {Delehaye}}, \bibinfo {author} {\bibfnamefont {S.}~\bibnamefont {Laurent}},
  \bibinfo {author} {\bibfnamefont {A.~T.}\ \bibnamefont {Grier}}, \bibinfo
  {author} {\bibfnamefont {M.}~\bibnamefont {Pierce}}, \bibinfo {author}
  {\bibfnamefont {B.~S.}\ \bibnamefont {Rem}}, \bibinfo {author} {\bibfnamefont
  {F.}~\bibnamefont {Chevy}}, \ and\ \bibinfo {author} {\bibfnamefont
  {C.}~\bibnamefont {Salomon}},\ }\bibfield  {title} {\enquote {\bibinfo
  {title} {A mixture of bose and fermi superfluids},}\ }\href {\doibase
  10.1126/science.1255380} {\bibfield  {journal} {\bibinfo  {journal}
  {Science}\ }\textbf {\bibinfo {volume} {345}},\ \bibinfo {pages} {1035--1038}
  (\bibinfo {year} {2014})},\ \Eprint
  {http://arxiv.org/abs/https://www.science.org/doi/pdf/10.1126/science.1255380}
  {https://www.science.org/doi/pdf/10.1126/science.1255380} \BibitemShut
  {NoStop}%
\bibitem [{\citenamefont {Wacker}\ \emph {et~al.}(2015)\citenamefont {Wacker},
  \citenamefont {J\o{}rgensen}, \citenamefont {Birkmose}, \citenamefont
  {Horchani}, \citenamefont {Ertmer}, \citenamefont {Klempt}, \citenamefont
  {Winter}, \citenamefont {Sherson},\ and\ \citenamefont {Arlt}}]{2015_Arlt}%
  \BibitemOpen
  \bibfield  {author} {\bibinfo {author} {\bibfnamefont {L.}~\bibnamefont
  {Wacker}}, \bibinfo {author} {\bibfnamefont {N.~B.}\ \bibnamefont
  {J\o{}rgensen}}, \bibinfo {author} {\bibfnamefont {D.}~\bibnamefont
  {Birkmose}}, \bibinfo {author} {\bibfnamefont {R.}~\bibnamefont {Horchani}},
  \bibinfo {author} {\bibfnamefont {W.}~\bibnamefont {Ertmer}}, \bibinfo
  {author} {\bibfnamefont {C.}~\bibnamefont {Klempt}}, \bibinfo {author}
  {\bibfnamefont {N.}~\bibnamefont {Winter}}, \bibinfo {author} {\bibfnamefont
  {J.}~\bibnamefont {Sherson}}, \ and\ \bibinfo {author} {\bibfnamefont
  {J.~J.}\ \bibnamefont {Arlt}},\ }\bibfield  {title} {\enquote {\bibinfo
  {title} {Tunable dual-species bose-einstein condensates of $^{39}\mathrm{K}$
  and $^{87}\mathrm{Rb}$},}\ }\href {\doibase 10.1103/PhysRevA.92.053602}
  {\bibfield  {journal} {\bibinfo  {journal} {Phys. Rev. A}\ }\textbf {\bibinfo
  {volume} {92}},\ \bibinfo {pages} {053602} (\bibinfo {year}
  {2015})}\BibitemShut {NoStop}%
\bibitem [{\citenamefont {Ravensbergen}\ \emph {et~al.}(2018)\citenamefont
  {Ravensbergen}, \citenamefont {Corre}, \citenamefont {Soave}, \citenamefont
  {Kreyer}, \citenamefont {Kirilov},\ and\ \citenamefont {Grimm}}]{2018_Grimm}%
  \BibitemOpen
  \bibfield  {author} {\bibinfo {author} {\bibfnamefont {C.}~\bibnamefont
  {Ravensbergen}}, \bibinfo {author} {\bibfnamefont {V.}~\bibnamefont {Corre}},
  \bibinfo {author} {\bibfnamefont {E.}~\bibnamefont {Soave}}, \bibinfo
  {author} {\bibfnamefont {M.}~\bibnamefont {Kreyer}}, \bibinfo {author}
  {\bibfnamefont {E.}~\bibnamefont {Kirilov}}, \ and\ \bibinfo {author}
  {\bibfnamefont {R.}~\bibnamefont {Grimm}},\ }\bibfield  {title} {\enquote
  {\bibinfo {title} {Production of a degenerate fermi-fermi mixture of
  dysprosium and potassium atoms},}\ }\href {\doibase
  10.1103/PhysRevA.98.063624} {\bibfield  {journal} {\bibinfo  {journal} {Phys.
  Rev. A}\ }\textbf {\bibinfo {volume} {98}},\ \bibinfo {pages} {063624}
  (\bibinfo {year} {2018})}\BibitemShut {NoStop}%
\bibitem [{\citenamefont {Castilho}\ \emph {et~al.}(2019)\citenamefont
  {Castilho}, \citenamefont {Pedrozo-Pe{\~n}afiel}, \citenamefont {Gutierrez},
  \citenamefont {Mazo}, \citenamefont {Roati}, \citenamefont {Farias},\ and\
  \citenamefont {Bagnato}}]{Castilho_2019}%
  \BibitemOpen
  \bibfield  {author} {\bibinfo {author} {\bibfnamefont {P.~C.~M.}\
  \bibnamefont {Castilho}}, \bibinfo {author} {\bibfnamefont {E.}~\bibnamefont
  {Pedrozo-Pe{\~n}afiel}}, \bibinfo {author} {\bibfnamefont {E.~M.}\
  \bibnamefont {Gutierrez}}, \bibinfo {author} {\bibfnamefont {P.~L.}\
  \bibnamefont {Mazo}}, \bibinfo {author} {\bibfnamefont {G.}~\bibnamefont
  {Roati}}, \bibinfo {author} {\bibfnamefont {K.~M.}\ \bibnamefont {Farias}}, \
  and\ \bibinfo {author} {\bibfnamefont {V.~S.}\ \bibnamefont {Bagnato}},\
  }\bibfield  {title} {\enquote {\bibinfo {title} {A compact experimental
  machine for studying tunable bose--bose superfluid mixtures},}\ }\href
  {\doibase 10.1088/1612-202X/ab00fb} {\bibfield  {journal} {\bibinfo
  {journal} {Laser Physics Letters}\ }\textbf {\bibinfo {volume} {16}},\
  \bibinfo {pages} {035501} (\bibinfo {year} {2019})}\BibitemShut {NoStop}%
\bibitem [{\citenamefont {Chikkatur}\ \emph {et~al.}(2000)\citenamefont
  {Chikkatur}, \citenamefont {G\"orlitz}, \citenamefont {Stamper-Kurn},
  \citenamefont {Inouye}, \citenamefont {Gupta},\ and\ \citenamefont
  {Ketterle}}]{ChikkaturPRL2000}%
  \BibitemOpen
  \bibfield  {author} {\bibinfo {author} {\bibfnamefont {A.~P.}\ \bibnamefont
  {Chikkatur}}, \bibinfo {author} {\bibfnamefont {A.}~\bibnamefont
  {G\"orlitz}}, \bibinfo {author} {\bibfnamefont {D.~M.}\ \bibnamefont
  {Stamper-Kurn}}, \bibinfo {author} {\bibfnamefont {S.}~\bibnamefont
  {Inouye}}, \bibinfo {author} {\bibfnamefont {S.}~\bibnamefont {Gupta}}, \
  and\ \bibinfo {author} {\bibfnamefont {W.}~\bibnamefont {Ketterle}},\
  }\bibfield  {title} {\enquote {\bibinfo {title} {Suppression and enhancement
  of impurity scattering in a bose-einstein condensate},}\ }\href {\doibase
  10.1103/PhysRevLett.85.483} {\bibfield  {journal} {\bibinfo  {journal} {Phys.
  Rev. Lett.}\ }\textbf {\bibinfo {volume} {85}},\ \bibinfo {pages} {483--486}
  (\bibinfo {year} {2000})}\BibitemShut {NoStop}%
\bibitem [{\citenamefont {Ospelkaus}\ \emph
  {et~al.}(2006{\natexlab{b}})\citenamefont {Ospelkaus}, \citenamefont
  {Ospelkaus}, \citenamefont {Wille}, \citenamefont {Succo}, \citenamefont
  {Ernst}, \citenamefont {Sengstock},\ and\ \citenamefont
  {Bongs}}]{OspelkausPRL2006}%
  \BibitemOpen
  \bibfield  {author} {\bibinfo {author} {\bibfnamefont {S.}~\bibnamefont
  {Ospelkaus}}, \bibinfo {author} {\bibfnamefont {C.}~\bibnamefont
  {Ospelkaus}}, \bibinfo {author} {\bibfnamefont {O.}~\bibnamefont {Wille}},
  \bibinfo {author} {\bibfnamefont {M.}~\bibnamefont {Succo}}, \bibinfo
  {author} {\bibfnamefont {P.}~\bibnamefont {Ernst}}, \bibinfo {author}
  {\bibfnamefont {K.}~\bibnamefont {Sengstock}}, \ and\ \bibinfo {author}
  {\bibfnamefont {K.}~\bibnamefont {Bongs}},\ }\bibfield  {title} {\enquote
  {\bibinfo {title} {Localization of bosonic atoms by fermionic impurities in a
  three-dimensional optical lattice},}\ }\href {\doibase
  10.1103/PhysRevLett.96.180403} {\bibfield  {journal} {\bibinfo  {journal}
  {Phys. Rev. Lett.}\ }\textbf {\bibinfo {volume} {96}},\ \bibinfo {pages}
  {180403} (\bibinfo {year} {2006}{\natexlab{b}})}\BibitemShut {NoStop}%
\bibitem [{\citenamefont {Bruderer}\ \emph {et~al.}(2007)\citenamefont
  {Bruderer}, \citenamefont {Klein}, \citenamefont {Clark},\ and\ \citenamefont
  {Jaksch}}]{BrudererELP2008}%
  \BibitemOpen
  \bibfield  {author} {\bibinfo {author} {\bibfnamefont {M.}~\bibnamefont
  {Bruderer}}, \bibinfo {author} {\bibfnamefont {A.}~\bibnamefont {Klein}},
  \bibinfo {author} {\bibfnamefont {S.}~\bibnamefont {Clark}}, \ and\ \bibinfo
  {author} {\bibfnamefont {D.}~\bibnamefont {Jaksch}},\ }\bibfield  {title}
  {\enquote {\bibinfo {title} {Transport of strong-coupling polarons in optical
  lattices},}\ }\href {\doibase 10.1088/1367-2630/10/3/033015} {\bibfield
  {journal} {\bibinfo  {journal} {New Journal of Physics}\ }\textbf {\bibinfo
  {volume} {10}},\ \bibinfo {pages} {033015} (\bibinfo {year}
  {2007})}\BibitemShut {NoStop}%
\bibitem [{\citenamefont {Petrov}(2015)}]{PetrovPRL2015}%
  \BibitemOpen
  \bibfield  {author} {\bibinfo {author} {\bibfnamefont {D.~S.}\ \bibnamefont
  {Petrov}},\ }\bibfield  {title} {\enquote {\bibinfo {title} {Quantum
  mechanical stabilization of a collapsing bose-bose mixture},}\ }\href
  {\doibase 10.1103/PhysRevLett.115.155302} {\bibfield  {journal} {\bibinfo
  {journal} {Phys. Rev. Lett.}\ }\textbf {\bibinfo {volume} {115}},\ \bibinfo
  {pages} {155302} (\bibinfo {year} {2015})}\BibitemShut {NoStop}%
\bibitem [{\citenamefont {Cabrera}\ \emph {et~al.}(2018)\citenamefont
  {Cabrera}, \citenamefont {Tanzi}, \citenamefont {Sanz}, \citenamefont
  {Naylor}, \citenamefont {Thomas}, \citenamefont {Cheiney},\ and\
  \citenamefont {Tarruell}}]{CabreraScience2018}%
  \BibitemOpen
  \bibfield  {author} {\bibinfo {author} {\bibfnamefont {C.~R.}\ \bibnamefont
  {Cabrera}}, \bibinfo {author} {\bibfnamefont {L.}~\bibnamefont {Tanzi}},
  \bibinfo {author} {\bibfnamefont {J.}~\bibnamefont {Sanz}}, \bibinfo {author}
  {\bibfnamefont {B.}~\bibnamefont {Naylor}}, \bibinfo {author} {\bibfnamefont
  {P.}~\bibnamefont {Thomas}}, \bibinfo {author} {\bibfnamefont
  {P.}~\bibnamefont {Cheiney}}, \ and\ \bibinfo {author} {\bibfnamefont
  {L.}~\bibnamefont {Tarruell}},\ }\bibfield  {title} {\enquote {\bibinfo
  {title} {Quantum liquid droplets in a mixture of bose-einstein
  condensates},}\ }\href {\doibase 10.1126/science.aao5686} {\bibfield
  {journal} {\bibinfo  {journal} {Science}\ }\textbf {\bibinfo {volume}
  {359}},\ \bibinfo {pages} {301--304} (\bibinfo {year} {2018})},\ \Eprint
  {http://arxiv.org/abs/https://www.science.org/doi/pdf/10.1126/science.aao5686}
  {https://www.science.org/doi/pdf/10.1126/science.aao5686} \BibitemShut
  {NoStop}%
\bibitem [{\citenamefont {Semeghini}\ \emph {et~al.}(2018)\citenamefont
  {Semeghini}, \citenamefont {Ferioli}, \citenamefont {Masi}, \citenamefont
  {Mazzinghi}, \citenamefont {Wolswijk}, \citenamefont {Minardi}, \citenamefont
  {Modugno}, \citenamefont {Modugno}, \citenamefont {Inguscio},\ and\
  \citenamefont {Fattori}}]{SemeghiniPRL2018}%
  \BibitemOpen
  \bibfield  {author} {\bibinfo {author} {\bibfnamefont {G.}~\bibnamefont
  {Semeghini}}, \bibinfo {author} {\bibfnamefont {G.}~\bibnamefont {Ferioli}},
  \bibinfo {author} {\bibfnamefont {L.}~\bibnamefont {Masi}}, \bibinfo {author}
  {\bibfnamefont {C.}~\bibnamefont {Mazzinghi}}, \bibinfo {author}
  {\bibfnamefont {L.}~\bibnamefont {Wolswijk}}, \bibinfo {author}
  {\bibfnamefont {F.}~\bibnamefont {Minardi}}, \bibinfo {author} {\bibfnamefont
  {M.}~\bibnamefont {Modugno}}, \bibinfo {author} {\bibfnamefont
  {G.}~\bibnamefont {Modugno}}, \bibinfo {author} {\bibfnamefont
  {M.}~\bibnamefont {Inguscio}}, \ and\ \bibinfo {author} {\bibfnamefont
  {M.}~\bibnamefont {Fattori}},\ }\bibfield  {title} {\enquote {\bibinfo
  {title} {Self-bound quantum droplets of atomic mixtures in free space},}\
  }\href {\doibase 10.1103/PhysRevLett.120.235301} {\bibfield  {journal}
  {\bibinfo  {journal} {Phys. Rev. Lett.}\ }\textbf {\bibinfo {volume} {120}},\
  \bibinfo {pages} {235301} (\bibinfo {year} {2018})}\BibitemShut {NoStop}%
\bibitem [{\citenamefont {D'Errico}\ \emph {et~al.}(2019)\citenamefont
  {D'Errico}, \citenamefont {Burchianti}, \citenamefont {Prevedelli},
  \citenamefont {Salasnich}, \citenamefont {Ancilotto}, \citenamefont
  {Modugno}, \citenamefont {Minardi},\ and\ \citenamefont
  {Fort}}]{2019_Chiara}%
  \BibitemOpen
  \bibfield  {author} {\bibinfo {author} {\bibfnamefont {C.}~\bibnamefont
  {D'Errico}}, \bibinfo {author} {\bibfnamefont {A.}~\bibnamefont
  {Burchianti}}, \bibinfo {author} {\bibfnamefont {M.}~\bibnamefont
  {Prevedelli}}, \bibinfo {author} {\bibfnamefont {L.}~\bibnamefont
  {Salasnich}}, \bibinfo {author} {\bibfnamefont {F.}~\bibnamefont
  {Ancilotto}}, \bibinfo {author} {\bibfnamefont {M.}~\bibnamefont {Modugno}},
  \bibinfo {author} {\bibfnamefont {F.}~\bibnamefont {Minardi}}, \ and\
  \bibinfo {author} {\bibfnamefont {C.}~\bibnamefont {Fort}},\ }\bibfield
  {title} {\enquote {\bibinfo {title} {Observation of quantum droplets in a
  heteronuclear bosonic mixture},}\ }\href {\doibase
  10.1103/PhysRevResearch.1.033155} {\bibfield  {journal} {\bibinfo  {journal}
  {Phys. Rev. Research}\ }\textbf {\bibinfo {volume} {1}},\ \bibinfo {pages}
  {033155} (\bibinfo {year} {2019})}\BibitemShut {NoStop}%
\bibitem [{\citenamefont {\ifmmode~\dot{Z}\else \.{Z}\fi{}uchowski}\ and\
  \citenamefont {Hutson}(2010)}]{2010PRAHutson}%
  \BibitemOpen
  \bibfield  {author} {\bibinfo {author} {\bibfnamefont {P.~S.}\ \bibnamefont
  {\ifmmode~\dot{Z}\else \.{Z}\fi{}uchowski}}\ and\ \bibinfo {author}
  {\bibfnamefont {J.~M.}\ \bibnamefont {Hutson}},\ }\bibfield  {title}
  {\enquote {\bibinfo {title} {Reactions of ultracold alkali-metal dimers},}\
  }\href {\doibase 10.1103/PhysRevA.81.060703} {\bibfield  {journal} {\bibinfo
  {journal} {Phys. Rev. A}\ }\textbf {\bibinfo {volume} {81}},\ \bibinfo
  {pages} {060703} (\bibinfo {year} {2010})}\BibitemShut {NoStop}%
\bibitem [{\citenamefont {Voges}\ \emph {et~al.}(2020)\citenamefont {Voges},
  \citenamefont {Gersema}, \citenamefont {Meyer~zum Alten~Borgloh},
  \citenamefont {Schulze}, \citenamefont {Hartmann}, \citenamefont {Zenesini},\
  and\ \citenamefont {Ospelkaus}}]{VogesPRl2019}%
  \BibitemOpen
  \bibfield  {author} {\bibinfo {author} {\bibfnamefont {K.~K.}\ \bibnamefont
  {Voges}}, \bibinfo {author} {\bibfnamefont {P.}~\bibnamefont {Gersema}},
  \bibinfo {author} {\bibfnamefont {M.}~\bibnamefont {Meyer~zum
  Alten~Borgloh}}, \bibinfo {author} {\bibfnamefont {T.~A.}\ \bibnamefont
  {Schulze}}, \bibinfo {author} {\bibfnamefont {T.}~\bibnamefont {Hartmann}},
  \bibinfo {author} {\bibfnamefont {A.}~\bibnamefont {Zenesini}}, \ and\
  \bibinfo {author} {\bibfnamefont {S.}~\bibnamefont {Ospelkaus}},\ }\bibfield
  {title} {\enquote {\bibinfo {title} {Ultracold gas of bosonic
  $^{23}\mathrm{Na}^{39}\mathrm{K}$ ground-state molecules},}\ }\href {\doibase
  10.1103/PhysRevLett.125.083401} {\bibfield  {journal} {\bibinfo  {journal}
  {Phys. Rev. Lett.}\ }\textbf {\bibinfo {volume} {125}},\ \bibinfo {pages}
  {083401} (\bibinfo {year} {2020})}\BibitemShut {NoStop}%
\bibitem [{\citenamefont {Park}, \citenamefont {Will},\ and\ \citenamefont
  {Zwierlein}(2015)}]{ParkPRL2015}%
  \BibitemOpen
  \bibfield  {author} {\bibinfo {author} {\bibfnamefont {J.}~\bibnamefont
  {Park}}, \bibinfo {author} {\bibfnamefont {S.}~\bibnamefont {Will}}, \ and\
  \bibinfo {author} {\bibfnamefont {M.}~\bibnamefont {Zwierlein}},\ }\bibfield
  {title} {\enquote {\bibinfo {title} {Ultracold dipolar gas of fermionic na 23
  k 40 molecules in their absolute ground state},}\ }\href {\doibase
  10.1103/PhysRevLett.114.205302} {\bibfield  {journal} {\bibinfo  {journal}
  {Physical review letters}\ }\textbf {\bibinfo {volume} {114}} (\bibinfo
  {year} {2015}),\ 10.1103/PhysRevLett.114.205302}\BibitemShut {NoStop}%
\bibitem [{\citenamefont {Ospelkaus}\ \emph {et~al.}(2010)\citenamefont
  {Ospelkaus}, \citenamefont {Ni}, \citenamefont {Wang}, \citenamefont
  {de~Miranda}, \citenamefont {Neyenhuis}, \citenamefont {Qu{\'e}m{\'e}ner},
  \citenamefont {Julienne}, \citenamefont {Bohn}, \citenamefont {Jin},\ and\
  \citenamefont {Ye}}]{OspelkausScience2010}%
  \BibitemOpen
  \bibfield  {author} {\bibinfo {author} {\bibfnamefont {S.}~\bibnamefont
  {Ospelkaus}}, \bibinfo {author} {\bibfnamefont {K.-K.}\ \bibnamefont {Ni}},
  \bibinfo {author} {\bibfnamefont {D.}~\bibnamefont {Wang}}, \bibinfo {author}
  {\bibfnamefont {M.~H.~G.}\ \bibnamefont {de~Miranda}}, \bibinfo {author}
  {\bibfnamefont {B.}~\bibnamefont {Neyenhuis}}, \bibinfo {author}
  {\bibfnamefont {G.}~\bibnamefont {Qu{\'e}m{\'e}ner}}, \bibinfo {author}
  {\bibfnamefont {P.~S.}\ \bibnamefont {Julienne}}, \bibinfo {author}
  {\bibfnamefont {J.~L.}\ \bibnamefont {Bohn}}, \bibinfo {author}
  {\bibfnamefont {D.~S.}\ \bibnamefont {Jin}}, \ and\ \bibinfo {author}
  {\bibfnamefont {J.}~\bibnamefont {Ye}},\ }\bibfield  {title} {\enquote
  {\bibinfo {title} {Quantum-state controlled chemical reactions of ultracold
  potassium-rubidium molecules},}\ }\href {\doibase 10.1126/science.1184121}
  {\bibfield  {journal} {\bibinfo  {journal} {Science}\ }\textbf {\bibinfo
  {volume} {327}},\ \bibinfo {pages} {853--857} (\bibinfo {year} {2010})},\
  \Eprint
  {http://arxiv.org/abs/https://www.science.org/doi/pdf/10.1126/science.1184121}
  {https://www.science.org/doi/pdf/10.1126/science.1184121} \BibitemShut
  {NoStop}%
\bibitem [{\citenamefont {Marco}\ \emph {et~al.}(2019)\citenamefont {Marco},
  \citenamefont {Valtolina}, \citenamefont {Matsuda}, \citenamefont {Tobias},
  \citenamefont {Covey},\ and\ \citenamefont {Ye}}]{DeMarcoScience2019}%
  \BibitemOpen
  \bibfield  {author} {\bibinfo {author} {\bibfnamefont {L.~D.}\ \bibnamefont
  {Marco}}, \bibinfo {author} {\bibfnamefont {G.}~\bibnamefont {Valtolina}},
  \bibinfo {author} {\bibfnamefont {K.}~\bibnamefont {Matsuda}}, \bibinfo
  {author} {\bibfnamefont {W.~G.}\ \bibnamefont {Tobias}}, \bibinfo {author}
  {\bibfnamefont {J.~P.}\ \bibnamefont {Covey}}, \ and\ \bibinfo {author}
  {\bibfnamefont {J.}~\bibnamefont {Ye}},\ }\bibfield  {title} {\enquote
  {\bibinfo {title} {A degenerate fermi gas of polar molecules},}\ }\href
  {\doibase 10.1126/science.aau7230} {\bibfield  {journal} {\bibinfo  {journal}
  {Science}\ }\textbf {\bibinfo {volume} {363}},\ \bibinfo {pages} {853--856}
  (\bibinfo {year} {2019})},\ \Eprint
  {http://arxiv.org/abs/https://www.science.org/doi/pdf/10.1126/science.aau7230}
  {https://www.science.org/doi/pdf/10.1126/science.aau7230} \BibitemShut
  {NoStop}%
\bibitem [{\citenamefont {Takekoshi}\ \emph {et~al.}(2014)\citenamefont
  {Takekoshi}, \citenamefont {Reichs\"ollner}, \citenamefont {Schindewolf},
  \citenamefont {Hutson}, \citenamefont {Le~Sueur}, \citenamefont {Dulieu},
  \citenamefont {Ferlaino}, \citenamefont {Grimm},\ and\ \citenamefont
  {N\"agerl}}]{TakekoshiPRL2014}%
  \BibitemOpen
  \bibfield  {author} {\bibinfo {author} {\bibfnamefont {T.}~\bibnamefont
  {Takekoshi}}, \bibinfo {author} {\bibfnamefont {L.}~\bibnamefont
  {Reichs\"ollner}}, \bibinfo {author} {\bibfnamefont {A.}~\bibnamefont
  {Schindewolf}}, \bibinfo {author} {\bibfnamefont {J.~M.}\ \bibnamefont
  {Hutson}}, \bibinfo {author} {\bibfnamefont {C.~R.}\ \bibnamefont
  {Le~Sueur}}, \bibinfo {author} {\bibfnamefont {O.}~\bibnamefont {Dulieu}},
  \bibinfo {author} {\bibfnamefont {F.}~\bibnamefont {Ferlaino}}, \bibinfo
  {author} {\bibfnamefont {R.}~\bibnamefont {Grimm}}, \ and\ \bibinfo {author}
  {\bibfnamefont {H.-C.}\ \bibnamefont {N\"agerl}},\ }\bibfield  {title}
  {\enquote {\bibinfo {title} {Ultracold dense samples of dipolar rbcs
  molecules in the rovibrational and hyperfine ground state},}\ }\href
  {\doibase 10.1103/PhysRevLett.113.205301} {\bibfield  {journal} {\bibinfo
  {journal} {Phys. Rev. Lett.}\ }\textbf {\bibinfo {volume} {113}},\ \bibinfo
  {pages} {205301} (\bibinfo {year} {2014})}\BibitemShut {NoStop}%
\bibitem [{\citenamefont {Damski}\ \emph {et~al.}(2003)\citenamefont {Damski},
  \citenamefont {Santos}, \citenamefont {Tiemann}, \citenamefont {Lewenstein},
  \citenamefont {Kotochigova}, \citenamefont {Julienne},\ and\ \citenamefont
  {Zoller}}]{Damski_PRL_2003}%
  \BibitemOpen
  \bibfield  {author} {\bibinfo {author} {\bibfnamefont {B.}~\bibnamefont
  {Damski}}, \bibinfo {author} {\bibfnamefont {L.}~\bibnamefont {Santos}},
  \bibinfo {author} {\bibfnamefont {E.}~\bibnamefont {Tiemann}}, \bibinfo
  {author} {\bibfnamefont {M.}~\bibnamefont {Lewenstein}}, \bibinfo {author}
  {\bibfnamefont {S.}~\bibnamefont {Kotochigova}}, \bibinfo {author}
  {\bibfnamefont {P.}~\bibnamefont {Julienne}}, \ and\ \bibinfo {author}
  {\bibfnamefont {P.}~\bibnamefont {Zoller}},\ }\bibfield  {title} {\enquote
  {\bibinfo {title} {Creation of a dipolar superfluid in optical lattices},}\
  }\href {\doibase 10.1103/PhysRevLett.90.110401} {\bibfield  {journal}
  {\bibinfo  {journal} {Phys. Rev. Lett.}\ }\textbf {\bibinfo {volume} {90}},\
  \bibinfo {pages} {110401} (\bibinfo {year} {2003})}\BibitemShut {NoStop}%
\bibitem [{\citenamefont {Lahaye}\ \emph {et~al.}(2009)\citenamefont {Lahaye},
  \citenamefont {Menotti}, \citenamefont {Santos}, \citenamefont {Lewenstein},\
  and\ \citenamefont {Pfau}}]{Lahaye_2009}%
  \BibitemOpen
  \bibfield  {author} {\bibinfo {author} {\bibfnamefont {T.}~\bibnamefont
  {Lahaye}}, \bibinfo {author} {\bibfnamefont {C.}~\bibnamefont {Menotti}},
  \bibinfo {author} {\bibfnamefont {L.}~\bibnamefont {Santos}}, \bibinfo
  {author} {\bibfnamefont {M.}~\bibnamefont {Lewenstein}}, \ and\ \bibinfo
  {author} {\bibfnamefont {T.}~\bibnamefont {Pfau}},\ }\bibfield  {title}
  {\enquote {\bibinfo {title} {The physics of dipolar bosonic quantum gases},}\
  }\href {\doibase 10.1088/0034-4885/72/12/126401} {\bibfield  {journal}
  {\bibinfo  {journal} {Reports on Progress in Physics}\ }\textbf {\bibinfo
  {volume} {72}},\ \bibinfo {pages} {126401} (\bibinfo {year}
  {2009})}\BibitemShut {NoStop}%
\bibitem [{\citenamefont {Schindewolf}\ \emph {et~al.}(2022)\citenamefont
  {Schindewolf}, \citenamefont {Bause}, \citenamefont {Chen}, \citenamefont
  {Duda}, \citenamefont {Karman}, \citenamefont {Bloch},\ and\ \citenamefont
  {Luo}}]{2022_Bloch}%
  \BibitemOpen
  \bibfield  {author} {\bibinfo {author} {\bibfnamefont {A.}~\bibnamefont
  {Schindewolf}}, \bibinfo {author} {\bibfnamefont {R.}~\bibnamefont {Bause}},
  \bibinfo {author} {\bibfnamefont {X.-Y.}\ \bibnamefont {Chen}}, \bibinfo
  {author} {\bibfnamefont {M.}~\bibnamefont {Duda}}, \bibinfo {author}
  {\bibfnamefont {T.}~\bibnamefont {Karman}}, \bibinfo {author} {\bibfnamefont
  {I.}~\bibnamefont {Bloch}}, \ and\ \bibinfo {author} {\bibfnamefont {X.-Y.}\
  \bibnamefont {Luo}},\ }\bibfield  {title} {\enquote {\bibinfo {title}
  {Evaporation of microwave-shielded polar molecules to quantum degeneracy},}\
  }\href {\doibase 10.1038/s41586-022-04900-0} {\bibfield  {journal} {\bibinfo
  {journal} {Nature}\ }\textbf {\bibinfo {volume} {607}},\ \bibinfo {pages}
  {677--681} (\bibinfo {year} {2022})}\BibitemShut {NoStop}%
\bibitem [{\citenamefont {Will}\ \emph {et~al.}(2016)\citenamefont {Will},
  \citenamefont {Park}, \citenamefont {Yan}, \citenamefont {Loh},\ and\
  \citenamefont {Zwierlein}}]{2016_Zwerlein}%
  \BibitemOpen
  \bibfield  {author} {\bibinfo {author} {\bibfnamefont {S.~A.}\ \bibnamefont
  {Will}}, \bibinfo {author} {\bibfnamefont {J.~W.}\ \bibnamefont {Park}},
  \bibinfo {author} {\bibfnamefont {Z.~Z.}\ \bibnamefont {Yan}}, \bibinfo
  {author} {\bibfnamefont {H.}~\bibnamefont {Loh}}, \ and\ \bibinfo {author}
  {\bibfnamefont {M.~W.}\ \bibnamefont {Zwierlein}},\ }\bibfield  {title}
  {\enquote {\bibinfo {title} {Coherent microwave control of ultracold
  $^{23}\mathrm{Na}^{40}\mathrm{K}$ molecules},}\ }\href {\doibase
  10.1103/PhysRevLett.116.225306} {\bibfield  {journal} {\bibinfo  {journal}
  {Phys. Rev. Lett.}\ }\textbf {\bibinfo {volume} {116}},\ \bibinfo {pages}
  {225306} (\bibinfo {year} {2016})}\BibitemShut {NoStop}%
\bibitem [{\citenamefont {Baranov}(2008)}]{2008_Baranov}%
  \BibitemOpen
  \bibfield  {author} {\bibinfo {author} {\bibfnamefont {M.}~\bibnamefont
  {Baranov}},\ }\bibfield  {title} {\enquote {\bibinfo {title} {Theoretical
  progress in many-body physics with ultracold dipolar gases},}\ }\href
  {\doibase https://doi.org/10.1016/j.physrep.2008.04.007} {\bibfield
  {journal} {\bibinfo  {journal} {Physics Reports}\ }\textbf {\bibinfo {volume}
  {464}},\ \bibinfo {pages} {71--111} (\bibinfo {year} {2008})}\BibitemShut
  {NoStop}%
\bibitem [{\citenamefont {Phillips}\ and\ \citenamefont
  {Metcalf}(1982)}]{PhillipsPRL1982}%
  \BibitemOpen
  \bibfield  {author} {\bibinfo {author} {\bibfnamefont {W.~D.}\ \bibnamefont
  {Phillips}}\ and\ \bibinfo {author} {\bibfnamefont {H.}~\bibnamefont
  {Metcalf}},\ }\bibfield  {title} {\enquote {\bibinfo {title} {Laser
  deceleration of an atomic beam},}\ }\href {\doibase
  10.1103/PhysRevLett.48.596} {\bibfield  {journal} {\bibinfo  {journal} {Phys.
  Rev. Lett.}\ }\textbf {\bibinfo {volume} {48}},\ \bibinfo {pages} {596--599}
  (\bibinfo {year} {1982})}\BibitemShut {NoStop}%
\bibitem [{\citenamefont {Lu}\ \emph {et~al.}(1996)\citenamefont {Lu},
  \citenamefont {Corwin}, \citenamefont {Renn}, \citenamefont {Anderson},
  \citenamefont {Cornell},\ and\ \citenamefont {Wieman}}]{1996PRLLu}%
  \BibitemOpen
  \bibfield  {author} {\bibinfo {author} {\bibfnamefont {Z.~T.}\ \bibnamefont
  {Lu}}, \bibinfo {author} {\bibfnamefont {K.~L.}\ \bibnamefont {Corwin}},
  \bibinfo {author} {\bibfnamefont {M.~J.}\ \bibnamefont {Renn}}, \bibinfo
  {author} {\bibfnamefont {M.~H.}\ \bibnamefont {Anderson}}, \bibinfo {author}
  {\bibfnamefont {E.~A.}\ \bibnamefont {Cornell}}, \ and\ \bibinfo {author}
  {\bibfnamefont {C.~E.}\ \bibnamefont {Wieman}},\ }\bibfield  {title}
  {\enquote {\bibinfo {title} {Low-velocity intense source of atoms from a
  magneto-optical trap},}\ }\href {\doibase 10.1103/PhysRevLett.77.3331}
  {\bibfield  {journal} {\bibinfo  {journal} {Phys. Rev. Lett.}\ }\textbf
  {\bibinfo {volume} {77}},\ \bibinfo {pages} {3331--3334} (\bibinfo {year}
  {1996})}\BibitemShut {NoStop}%
\bibitem [{\citenamefont {Schoser}\ \emph {et~al.}(2002)\citenamefont
  {Schoser}, \citenamefont {Bat\"ar}, \citenamefont {L\"ow}, \citenamefont
  {Schweikhard}, \citenamefont {Grabowski}, \citenamefont {Ovchinnikov},\ and\
  \citenamefont {Pfau}}]{2002PRASchoser}%
  \BibitemOpen
  \bibfield  {author} {\bibinfo {author} {\bibfnamefont {J.}~\bibnamefont
  {Schoser}}, \bibinfo {author} {\bibfnamefont {A.}~\bibnamefont {Bat\"ar}},
  \bibinfo {author} {\bibfnamefont {R.}~\bibnamefont {L\"ow}}, \bibinfo
  {author} {\bibfnamefont {V.}~\bibnamefont {Schweikhard}}, \bibinfo {author}
  {\bibfnamefont {A.}~\bibnamefont {Grabowski}}, \bibinfo {author}
  {\bibfnamefont {Y.~B.}\ \bibnamefont {Ovchinnikov}}, \ and\ \bibinfo {author}
  {\bibfnamefont {T.}~\bibnamefont {Pfau}},\ }\bibfield  {title} {\enquote
  {\bibinfo {title} {Intense source of cold rb atoms from a pure
  two-dimensional magneto-optical trap},}\ }\href {\doibase
  10.1103/PhysRevA.66.023410} {\bibfield  {journal} {\bibinfo  {journal} {Phys.
  Rev. A}\ }\textbf {\bibinfo {volume} {66}},\ \bibinfo {pages} {023410}
  (\bibinfo {year} {2002})}\BibitemShut {NoStop}%
\bibitem [{\citenamefont {Chaudhuri}, \citenamefont {Roy},\ and\ \citenamefont
  {Unnikrishnan}(2006)}]{2006_Chaudhuri}%
  \BibitemOpen
  \bibfield  {author} {\bibinfo {author} {\bibfnamefont {S.}~\bibnamefont
  {Chaudhuri}}, \bibinfo {author} {\bibfnamefont {S.}~\bibnamefont {Roy}}, \
  and\ \bibinfo {author} {\bibfnamefont {C.~S.}\ \bibnamefont {Unnikrishnan}},\
  }\bibfield  {title} {\enquote {\bibinfo {title} {Realization of an intense
  cold rb atomic beam based on a two-dimensional magneto-optical trap:
  Experiments and comparison with simulations},}\ }\href {\doibase
  10.1103/PhysRevA.74.023406} {\bibfield  {journal} {\bibinfo  {journal} {Phys.
  Rev. A}\ }\textbf {\bibinfo {volume} {74}},\ \bibinfo {pages} {023406}
  (\bibinfo {year} {2006})}\BibitemShut {NoStop}%
\bibitem [{\citenamefont {Dieckmann}\ \emph {et~al.}(1998)\citenamefont
  {Dieckmann}, \citenamefont {Spreeuw}, \citenamefont {Weidem\"uller},\ and\
  \citenamefont {Walraven}}]{1998PRADieckmann}%
  \BibitemOpen
  \bibfield  {author} {\bibinfo {author} {\bibfnamefont {K.}~\bibnamefont
  {Dieckmann}}, \bibinfo {author} {\bibfnamefont {R.~J.~C.}\ \bibnamefont
  {Spreeuw}}, \bibinfo {author} {\bibfnamefont {M.}~\bibnamefont
  {Weidem\"uller}}, \ and\ \bibinfo {author} {\bibfnamefont {J.~T.~M.}\
  \bibnamefont {Walraven}},\ }\bibfield  {title} {\enquote {\bibinfo {title}
  {Two-dimensional magneto-optical trap as a source of slow atoms},}\ }\href
  {\doibase 10.1103/PhysRevA.58.3891} {\bibfield  {journal} {\bibinfo
  {journal} {Phys. Rev. A}\ }\textbf {\bibinfo {volume} {58}},\ \bibinfo
  {pages} {3891--3895} (\bibinfo {year} {1998})}\BibitemShut {NoStop}%
\bibitem [{\citenamefont {Ravenhall}, \citenamefont {Yuen},\ and\ \citenamefont
  {Foot}(2021)}]{2021OERavenhall}%
  \BibitemOpen
  \bibfield  {author} {\bibinfo {author} {\bibfnamefont {S.}~\bibnamefont
  {Ravenhall}}, \bibinfo {author} {\bibfnamefont {B.}~\bibnamefont {Yuen}}, \
  and\ \bibinfo {author} {\bibfnamefont {C.}~\bibnamefont {Foot}},\ }\bibfield
  {title} {\enquote {\bibinfo {title} {High-flux, adjustable, compact cold-atom
  source},}\ }\href {\doibase 10.1364/OE.423662} {\bibfield  {journal}
  {\bibinfo  {journal} {Opt. Express}\ }\textbf {\bibinfo {volume} {29}},\
  \bibinfo {pages} {21143--21159} (\bibinfo {year} {2021})}\BibitemShut
  {NoStop}%
\bibitem [{\citenamefont {van~der Stam}\ \emph {et~al.}(2007)\citenamefont
  {van~der Stam}, \citenamefont {van Ooijen}, \citenamefont {Meppelink},
  \citenamefont {Vogels},\ and\ \citenamefont {van~der
  Straten}}]{Na_Zeemanslower_Straten}%
  \BibitemOpen
  \bibfield  {author} {\bibinfo {author} {\bibfnamefont {K.~M.~R.}\
  \bibnamefont {van~der Stam}}, \bibinfo {author} {\bibfnamefont {E.~D.}\
  \bibnamefont {van Ooijen}}, \bibinfo {author} {\bibfnamefont
  {R.}~\bibnamefont {Meppelink}}, \bibinfo {author} {\bibfnamefont {J.~M.}\
  \bibnamefont {Vogels}}, \ and\ \bibinfo {author} {\bibfnamefont
  {P.}~\bibnamefont {van~der Straten}},\ }\bibfield  {title} {\enquote
  {\bibinfo {title} {Large atom number bose-einstein condensate of sodium},}\
  }\href {\doibase 10.1063/1.2424439} {\bibfield  {journal} {\bibinfo
  {journal} {Review of Scientific Instruments}\ }\textbf {\bibinfo {volume}
  {78}},\ \bibinfo {pages} {013102} (\bibinfo {year} {2007})},\ \Eprint
  {http://arxiv.org/abs/https://doi.org/10.1063/1.2424439}
  {https://doi.org/10.1063/1.2424439} \BibitemShut {NoStop}%
\bibitem [{\citenamefont {Pedrozo-Pe{\~{n}}afiel}\ \emph
  {et~al.}(2016)\citenamefont {Pedrozo-Pe{\~{n}}afiel}, \citenamefont
  {Vivanco}, \citenamefont {Castilho}, \citenamefont {Paiva}, \citenamefont
  {Farias},\ and\ \citenamefont {Bagnato}}]{Pedrozo_Pe_afiel_2016}%
  \BibitemOpen
  \bibfield  {author} {\bibinfo {author} {\bibfnamefont {E.}~\bibnamefont
  {Pedrozo-Pe{\~{n}}afiel}}, \bibinfo {author} {\bibfnamefont {F.}~\bibnamefont
  {Vivanco}}, \bibinfo {author} {\bibfnamefont {P.}~\bibnamefont {Castilho}},
  \bibinfo {author} {\bibfnamefont {R.~R.}\ \bibnamefont {Paiva}}, \bibinfo
  {author} {\bibfnamefont {K.~M.}\ \bibnamefont {Farias}}, \ and\ \bibinfo
  {author} {\bibfnamefont {V.~S.}\ \bibnamefont {Bagnato}},\ }\bibfield
  {title} {\enquote {\bibinfo {title} {Direct comparison between a
  two-dimensional magneto-optical trap and a zeeman slower as sources of cold
  sodium atoms},}\ }\href {\doibase 10.1088/1612-2011/13/6/065501} {\bibfield
  {journal} {\bibinfo  {journal} {Laser Physics Letters}\ }\textbf {\bibinfo
  {volume} {13}},\ \bibinfo {pages} {065501} (\bibinfo {year}
  {2016})}\BibitemShut {NoStop}%
\bibitem [{\citenamefont {Catani}\ \emph {et~al.}(2006)\citenamefont {Catani},
  \citenamefont {Maioli}, \citenamefont {De~Sarlo}, \citenamefont {Minardi},\
  and\ \citenamefont {Inguscio}}]{CataniPRA2006}%
  \BibitemOpen
  \bibfield  {author} {\bibinfo {author} {\bibfnamefont {J.}~\bibnamefont
  {Catani}}, \bibinfo {author} {\bibfnamefont {P.}~\bibnamefont {Maioli}},
  \bibinfo {author} {\bibfnamefont {L.}~\bibnamefont {De~Sarlo}}, \bibinfo
  {author} {\bibfnamefont {F.}~\bibnamefont {Minardi}}, \ and\ \bibinfo
  {author} {\bibfnamefont {M.}~\bibnamefont {Inguscio}},\ }\bibfield  {title}
  {\enquote {\bibinfo {title} {Intense slow beams of bosonic potassium
  isotopes},}\ }\href {\doibase 10.1103/PhysRevA.73.033415} {\bibfield
  {journal} {\bibinfo  {journal} {Phys. Rev. A}\ }\textbf {\bibinfo {volume}
  {73}},\ \bibinfo {pages} {033415} (\bibinfo {year} {2006})}\BibitemShut
  {NoStop}%
\bibitem [{\citenamefont {Gr{\"o}bner}\ \emph {et~al.}(2016)\citenamefont
  {Gr{\"o}bner}, \citenamefont {Weinmann}, \citenamefont {Meinert},
  \citenamefont {Lauber}, \citenamefont {Kirilov},\ and\ \citenamefont
  {N{\"a}gerl}}]{GrobnerJMO2016}%
  \BibitemOpen
  \bibfield  {author} {\bibinfo {author} {\bibfnamefont {M.}~\bibnamefont
  {Gr{\"o}bner}}, \bibinfo {author} {\bibfnamefont {P.}~\bibnamefont
  {Weinmann}}, \bibinfo {author} {\bibfnamefont {F.}~\bibnamefont {Meinert}},
  \bibinfo {author} {\bibfnamefont {K.}~\bibnamefont {Lauber}}, \bibinfo
  {author} {\bibfnamefont {E.}~\bibnamefont {Kirilov}}, \ and\ \bibinfo
  {author} {\bibfnamefont {H.-C.}\ \bibnamefont {N{\"a}gerl}},\ }\bibfield
  {title} {\enquote {\bibinfo {title} {A new quantum gas apparatus for
  ultracold mixtures of k and cs and kcs ground-state molecules},}\ }\href
  {\doibase 10.1080/09500340.2016.1143051} {\bibfield  {journal} {\bibinfo
  {journal} {Journal of Modern Optics}\ }\textbf {\bibinfo {volume} {63}},\
  \bibinfo {pages} {1829--1839} (\bibinfo {year} {2016})},\ \Eprint
  {http://arxiv.org/abs/https://doi.org/10.1080/09500340.2016.1143051}
  {https://doi.org/10.1080/09500340.2016.1143051} \BibitemShut {NoStop}%
\bibitem [{\citenamefont {Lamporesi}\ \emph {et~al.}(2013)\citenamefont
  {Lamporesi}, \citenamefont {Donadello}, \citenamefont {Serafini},\ and\
  \citenamefont {Ferrari}}]{Ferrari}%
  \BibitemOpen
  \bibfield  {author} {\bibinfo {author} {\bibfnamefont {G.}~\bibnamefont
  {Lamporesi}}, \bibinfo {author} {\bibfnamefont {S.}~\bibnamefont
  {Donadello}}, \bibinfo {author} {\bibfnamefont {S.}~\bibnamefont {Serafini}},
  \ and\ \bibinfo {author} {\bibfnamefont {G.}~\bibnamefont {Ferrari}},\
  }\bibfield  {title} {\enquote {\bibinfo {title} {Compact high-flux source of
  cold sodium atoms},}\ }\href {\doibase 10.1063/1.4808375} {\bibfield
  {journal} {\bibinfo  {journal} {Review of Scientific Instruments}\ }\textbf
  {\bibinfo {volume} {84}},\ \bibinfo {pages} {063102} (\bibinfo {year}
  {2013})},\ \Eprint {http://arxiv.org/abs/https://doi.org/10.1063/1.4808375}
  {https://doi.org/10.1063/1.4808375} \BibitemShut {NoStop}%
\bibitem [{\citenamefont {Telles}\ \emph {et~al.}(2010)\citenamefont {Telles},
  \citenamefont {Ishikawa}, \citenamefont {Gibbs},\ and\ \citenamefont
  {Raman}}]{2010_Chandra_Raman}%
  \BibitemOpen
  \bibfield  {author} {\bibinfo {author} {\bibfnamefont {G.}~\bibnamefont
  {Telles}}, \bibinfo {author} {\bibfnamefont {T.}~\bibnamefont {Ishikawa}},
  \bibinfo {author} {\bibfnamefont {M.}~\bibnamefont {Gibbs}}, \ and\ \bibinfo
  {author} {\bibfnamefont {C.}~\bibnamefont {Raman}},\ }\bibfield  {title}
  {\enquote {\bibinfo {title} {Light-induced atomic desorption for loading a
  sodium magneto-optical trap},}\ }\href {\doibase 10.1103/PhysRevA.81.032710}
  {\bibfield  {journal} {\bibinfo  {journal} {Phys. Rev. A}\ }\textbf {\bibinfo
  {volume} {81}},\ \bibinfo {pages} {032710} (\bibinfo {year}
  {2010})}\BibitemShut {NoStop}%
\bibitem [{\citenamefont {Torralbo-Campo}\ \emph {et~al.}(2015)\citenamefont
  {Torralbo-Campo}, \citenamefont {Bruce}, \citenamefont {Smirne},\ and\
  \citenamefont {Cassettari}}]{2015_Campo}%
  \BibitemOpen
  \bibfield  {author} {\bibinfo {author} {\bibfnamefont {L.}~\bibnamefont
  {Torralbo-Campo}}, \bibinfo {author} {\bibfnamefont {G.~D.}\ \bibnamefont
  {Bruce}}, \bibinfo {author} {\bibfnamefont {G.}~\bibnamefont {Smirne}}, \
  and\ \bibinfo {author} {\bibfnamefont {D.}~\bibnamefont {Cassettari}},\
  }\bibfield  {title} {\enquote {\bibinfo {title} {Light-induced atomic
  desorption in a compact system for ultracold atoms},}\ }\href {\doibase
  10.1038/srep14729} {\bibfield  {journal} {\bibinfo  {journal} {Scientific
  Reports}\ }\textbf {\bibinfo {volume} {5}},\ \bibinfo {pages} {14729}
  (\bibinfo {year} {2015})}\BibitemShut {NoStop}%
\bibitem [{\citenamefont {Kwolek}\ \emph {et~al.}(2018)\citenamefont {Kwolek},
  \citenamefont {Goodman}, \citenamefont {Entner}, \citenamefont {Wells},
  \citenamefont {Narducci},\ and\ \citenamefont {Smith}}]{2018_Smith}%
  \BibitemOpen
  \bibfield  {author} {\bibinfo {author} {\bibfnamefont {J.~M.}\ \bibnamefont
  {Kwolek}}, \bibinfo {author} {\bibfnamefont {D.~S.}\ \bibnamefont {Goodman}},
  \bibinfo {author} {\bibfnamefont {S.~A.}\ \bibnamefont {Entner}}, \bibinfo
  {author} {\bibfnamefont {J.~E.}\ \bibnamefont {Wells}}, \bibinfo {author}
  {\bibfnamefont {F.~A.}\ \bibnamefont {Narducci}}, \ and\ \bibinfo {author}
  {\bibfnamefont {W.~W.}\ \bibnamefont {Smith}},\ }\bibfield  {title} {\enquote
  {\bibinfo {title} {Model-independent measurements of the sodium
  magneto-optical trap's excited-state population},}\ }\href {\doibase
  10.1103/PhysRevA.97.053420} {\bibfield  {journal} {\bibinfo  {journal} {Phys.
  Rev. A}\ }\textbf {\bibinfo {volume} {97}},\ \bibinfo {pages} {053420}
  (\bibinfo {year} {2018})}\BibitemShut {NoStop}%
\bibitem [{\citenamefont {Williamson}\ and\ \citenamefont
  {Walker}(1995)}]{1995_Walker}%
  \BibitemOpen
  \bibfield  {author} {\bibinfo {author} {\bibfnamefont {R.~S.}\ \bibnamefont
  {Williamson}}\ and\ \bibinfo {author} {\bibfnamefont {T.}~\bibnamefont
  {Walker}},\ }\bibfield  {title} {\enquote {\bibinfo {title} {Magneto-optical
  trapping and ultracold collisions of potassium atoms},}\ }\href {\doibase
  10.1364/JOSAB.12.001393} {\bibfield  {journal} {\bibinfo  {journal} {J. Opt.
  Soc. Am. B}\ }\textbf {\bibinfo {volume} {12}},\ \bibinfo {pages}
  {1393--1397} (\bibinfo {year} {1995})}\BibitemShut {NoStop}%
\bibitem [{\citenamefont {Metcalf}\ \emph {et~al.}(1999)\citenamefont
  {Metcalf}, \citenamefont {Van~der Straten}, \citenamefont {Birman},
  \citenamefont {Lynn},\ and\ \citenamefont {Stanley}}]{metcalf1999laser}%
  \BibitemOpen
  \bibfield  {author} {\bibinfo {author} {\bibfnamefont {H.}~\bibnamefont
  {Metcalf}}, \bibinfo {author} {\bibfnamefont {P.}~\bibnamefont {Van~der
  Straten}}, \bibinfo {author} {\bibfnamefont {J.}~\bibnamefont {Birman}},
  \bibinfo {author} {\bibfnamefont {J.}~\bibnamefont {Lynn}}, \ and\ \bibinfo
  {author} {\bibfnamefont {H.}~\bibnamefont {Stanley}},\ }\href
  {https://books.google.co.in/books?id=6p2ftAEACAAJ} {\emph {\bibinfo {title}
  {Laser Cooling and Trapping}}},\ Graduate texts in contemporary physics\
  (\bibinfo  {publisher} {Springer},\ \bibinfo {year} {1999})\BibitemShut
  {NoStop}%
\bibitem [{\citenamefont {Telles}\ \emph {et~al.}(2001)\citenamefont {Telles},
  \citenamefont {Garcia}, \citenamefont {Marcassa}, \citenamefont {Bagnato},
  \citenamefont {Ciampini}, \citenamefont {Fazzi}, \citenamefont {M\"uller},
  \citenamefont {Wilkowski},\ and\ \citenamefont {Arimondo}}]{2001_Telles}%
  \BibitemOpen
  \bibfield  {author} {\bibinfo {author} {\bibfnamefont {G.~D.}\ \bibnamefont
  {Telles}}, \bibinfo {author} {\bibfnamefont {W.}~\bibnamefont {Garcia}},
  \bibinfo {author} {\bibfnamefont {L.~G.}\ \bibnamefont {Marcassa}}, \bibinfo
  {author} {\bibfnamefont {V.~S.}\ \bibnamefont {Bagnato}}, \bibinfo {author}
  {\bibfnamefont {D.}~\bibnamefont {Ciampini}}, \bibinfo {author}
  {\bibfnamefont {M.}~\bibnamefont {Fazzi}}, \bibinfo {author} {\bibfnamefont
  {J.~H.}\ \bibnamefont {M\"uller}}, \bibinfo {author} {\bibfnamefont
  {D.}~\bibnamefont {Wilkowski}}, \ and\ \bibinfo {author} {\bibfnamefont
  {E.}~\bibnamefont {Arimondo}},\ }\bibfield  {title} {\enquote {\bibinfo
  {title} {Trap loss in a two-species rb-cs magneto-optical trap},}\ }\href
  {\doibase 10.1103/PhysRevA.63.033406} {\bibfield  {journal} {\bibinfo
  {journal} {Phys. Rev. A}\ }\textbf {\bibinfo {volume} {63}},\ \bibinfo
  {pages} {033406} (\bibinfo {year} {2001})}\BibitemShut {NoStop}%
\bibitem [{\citenamefont {Dinneen}\ \emph {et~al.}(1999)\citenamefont
  {Dinneen}, \citenamefont {Vogel}, \citenamefont {Arimondo}, \citenamefont
  {Hall},\ and\ \citenamefont {Gallagher}}]{1999_Arimondo}%
  \BibitemOpen
  \bibfield  {author} {\bibinfo {author} {\bibfnamefont {T.~P.}\ \bibnamefont
  {Dinneen}}, \bibinfo {author} {\bibfnamefont {K.~R.}\ \bibnamefont {Vogel}},
  \bibinfo {author} {\bibfnamefont {E.}~\bibnamefont {Arimondo}}, \bibinfo
  {author} {\bibfnamefont {J.~L.}\ \bibnamefont {Hall}}, \ and\ \bibinfo
  {author} {\bibfnamefont {A.}~\bibnamefont {Gallagher}},\ }\bibfield  {title}
  {\enquote {\bibinfo {title} {Cold collisions of
  ${\mathrm{sr}}^{*}\ensuremath{-}\mathrm{Sr}$ in a magneto-optical trap},}\
  }\href {\doibase 10.1103/PhysRevA.59.1216} {\bibfield  {journal} {\bibinfo
  {journal} {Phys. Rev. A}\ }\textbf {\bibinfo {volume} {59}},\ \bibinfo
  {pages} {1216--1222} (\bibinfo {year} {1999})}\BibitemShut {NoStop}%
\bibitem [{\citenamefont {Santos}\ \emph {et~al.}(1999)\citenamefont {Santos},
  \citenamefont {Nussenzveig}, \citenamefont {Antunes}, \citenamefont
  {Cardona},\ and\ \citenamefont {Bagnato}}]{1999_Santos}%
  \BibitemOpen
  \bibfield  {author} {\bibinfo {author} {\bibfnamefont {M.~S.}\ \bibnamefont
  {Santos}}, \bibinfo {author} {\bibfnamefont {P.}~\bibnamefont {Nussenzveig}},
  \bibinfo {author} {\bibfnamefont {A.}~\bibnamefont {Antunes}}, \bibinfo
  {author} {\bibfnamefont {P.~S.~P.}\ \bibnamefont {Cardona}}, \ and\ \bibinfo
  {author} {\bibfnamefont {V.~S.}\ \bibnamefont {Bagnato}},\ }\bibfield
  {title} {\enquote {\bibinfo {title} {Hyperfine-changing collision
  measurements in trap loss for mixed species in a magneto-optical trap},}\
  }\href {\doibase 10.1103/PhysRevA.60.3892} {\bibfield  {journal} {\bibinfo
  {journal} {Phys. Rev. A}\ }\textbf {\bibinfo {volume} {60}},\ \bibinfo
  {pages} {3892--3895} (\bibinfo {year} {1999})}\BibitemShut {NoStop}%
\bibitem [{\citenamefont {Chaudhuri}, \citenamefont {Roy},\ and\ \citenamefont
  {Unnikrishnan}(2008)}]{2008_CS}%
  \BibitemOpen
  \bibfield  {author} {\bibinfo {author} {\bibfnamefont {S.}~\bibnamefont
  {Chaudhuri}}, \bibinfo {author} {\bibfnamefont {S.}~\bibnamefont {Roy}}, \
  and\ \bibinfo {author} {\bibfnamefont {C.~S.}\ \bibnamefont {Unnikrishnan}},\
  }\bibfield  {title} {\enquote {\bibinfo {title} {Bose---einstein condensation
  in optical traps and in a 1d optical lattice},}\ }\href
  {http://www.jstor.org/stable/24102651} {\bibfield  {journal} {\bibinfo
  {journal} {Current Science}\ }\textbf {\bibinfo {volume} {95}},\ \bibinfo
  {pages} {1026--1034} (\bibinfo {year} {2008})}\BibitemShut {NoStop}%
\bibitem [{\citenamefont {Bhar}\ \emph {et~al.}(2022)\citenamefont {Bhar},
  \citenamefont {Swar}, \citenamefont {Satpathi}, \citenamefont {Sinha},
  \citenamefont {Sorkin}, \citenamefont {Chaudhuri},\ and\ \citenamefont
  {Roy}}]{2022_Response}%
  \BibitemOpen
  \bibfield  {author} {\bibinfo {author} {\bibfnamefont {S.}~\bibnamefont
  {Bhar}}, \bibinfo {author} {\bibfnamefont {M.}~\bibnamefont {Swar}}, \bibinfo
  {author} {\bibfnamefont {U.}~\bibnamefont {Satpathi}}, \bibinfo {author}
  {\bibfnamefont {S.}~\bibnamefont {Sinha}}, \bibinfo {author} {\bibfnamefont
  {R.}~\bibnamefont {Sorkin}}, \bibinfo {author} {\bibfnamefont
  {S.}~\bibnamefont {Chaudhuri}}, \ and\ \bibinfo {author} {\bibfnamefont
  {S.}~\bibnamefont {Roy}},\ }\bibfield  {title} {\enquote {\bibinfo {title}
  {Measurements and analysis of response function of cold atoms in optical
  molasses},}\ }\href {\doibase 10.1364/OPTCON.448535} {\bibfield  {journal}
  {\bibinfo  {journal} {Opt. Continuum}\ }\textbf {\bibinfo {volume} {1}},\
  \bibinfo {pages} {171--188} (\bibinfo {year} {2022})}\BibitemShut {NoStop}%
\bibitem [{\citenamefont {Swar}\ \emph {et~al.}(2021)\citenamefont {Swar},
  \citenamefont {Roy}, \citenamefont {Bhar}, \citenamefont {Roy},\ and\
  \citenamefont {Chaudhuri}}]{2021_Swar}%
  \BibitemOpen
  \bibfield  {author} {\bibinfo {author} {\bibfnamefont {M.}~\bibnamefont
  {Swar}}, \bibinfo {author} {\bibfnamefont {D.}~\bibnamefont {Roy}}, \bibinfo
  {author} {\bibfnamefont {S.}~\bibnamefont {Bhar}}, \bibinfo {author}
  {\bibfnamefont {S.}~\bibnamefont {Roy}}, \ and\ \bibinfo {author}
  {\bibfnamefont {S.}~\bibnamefont {Chaudhuri}},\ }\bibfield  {title} {\enquote
  {\bibinfo {title} {Detection of spin coherence in cold atoms via faraday
  rotation fluctuations},}\ }\href {\doibase 10.1103/PhysRevResearch.3.043171}
  {\bibfield  {journal} {\bibinfo  {journal} {Phys. Rev. Research}\ }\textbf
  {\bibinfo {volume} {3}},\ \bibinfo {pages} {043171} (\bibinfo {year}
  {2021})}\BibitemShut {NoStop}%
\bibitem [{\citenamefont {Oszmaniec}\ \emph {et~al.}(2016)\citenamefont
  {Oszmaniec}, \citenamefont {Augusiak}, \citenamefont {Gogolin}, \citenamefont
  {Ko\l{}ody\ifmmode~\acute{n}\else \'{n}\fi{}ski}, \citenamefont {Ac\'{\i}n},\
  and\ \citenamefont {Lewenstein}}]{2016_Lewenstein}%
  \BibitemOpen
  \bibfield  {author} {\bibinfo {author} {\bibfnamefont {M.}~\bibnamefont
  {Oszmaniec}}, \bibinfo {author} {\bibfnamefont {R.}~\bibnamefont {Augusiak}},
  \bibinfo {author} {\bibfnamefont {C.}~\bibnamefont {Gogolin}}, \bibinfo
  {author} {\bibfnamefont {J.}~\bibnamefont {Ko\l{}ody\ifmmode~\acute{n}\else
  \'{n}\fi{}ski}}, \bibinfo {author} {\bibfnamefont {A.}~\bibnamefont
  {Ac\'{\i}n}}, \ and\ \bibinfo {author} {\bibfnamefont {M.}~\bibnamefont
  {Lewenstein}},\ }\bibfield  {title} {\enquote {\bibinfo {title} {Random
  bosonic states for robust quantum metrology},}\ }\href {\doibase
  10.1103/PhysRevX.6.041044} {\bibfield  {journal} {\bibinfo  {journal} {Phys.
  Rev. X}\ }\textbf {\bibinfo {volume} {6}},\ \bibinfo {pages} {041044}
  (\bibinfo {year} {2016})}\BibitemShut {NoStop}%
\bibitem [{\citenamefont {Georgescu}, \citenamefont {Ashhab},\ and\
  \citenamefont {Nori}(2014)}]{2014_Nori}%
  \BibitemOpen
  \bibfield  {author} {\bibinfo {author} {\bibfnamefont {I.~M.}\ \bibnamefont
  {Georgescu}}, \bibinfo {author} {\bibfnamefont {S.}~\bibnamefont {Ashhab}}, \
  and\ \bibinfo {author} {\bibfnamefont {F.}~\bibnamefont {Nori}},\ }\bibfield
  {title} {\enquote {\bibinfo {title} {Quantum simulation},}\ }\href {\doibase
  10.1103/RevModPhys.86.153} {\bibfield  {journal} {\bibinfo  {journal} {Rev.
  Mod. Phys.}\ }\textbf {\bibinfo {volume} {86}},\ \bibinfo {pages} {153--185}
  (\bibinfo {year} {2014})}\BibitemShut {NoStop}%
\bibitem [{\citenamefont {Pelegr{\'\i}}, \citenamefont {Mompart},\ and\
  \citenamefont {Ahufinger}(2018)}]{2018_Pelegri}%
  \BibitemOpen
  \bibfield  {author} {\bibinfo {author} {\bibfnamefont {G.}~\bibnamefont
  {Pelegr{\'\i}}}, \bibinfo {author} {\bibfnamefont {J.}~\bibnamefont
  {Mompart}}, \ and\ \bibinfo {author} {\bibfnamefont {V.}~\bibnamefont
  {Ahufinger}},\ }\bibfield  {title} {\enquote {\bibinfo {title} {Quantum
  sensing using imbalanced counter-rotating bose--einstein condensate modes},}\
  }\href {\doibase 10.1088/1367-2630/aae107} {\bibfield  {journal} {\bibinfo
  {journal} {New Journal of Physics}\ }\textbf {\bibinfo {volume} {20}},\
  \bibinfo {pages} {103001} (\bibinfo {year} {2018})}\BibitemShut {NoStop}%
\end{thebibliography}%

\end{document}